\documentclass[twocolumn]{aastex63}

\usepackage{tikz}
\usepackage[english]{babel}
\usepackage[utf8]{inputenc}
\usepackage[colorinlistoftodos, color=green!40, prependcaption]{todonotes}
\usepackage{subfigure}
\usepackage{natbib}

\newcommand{\FMaxTeV}{F_{\rm TeV,\, Max}}
\newcommand{\YTeV}{Y_{\rm TeV}}
\newcommand{\LTeV}{L_{\rm TeV}}
\newcommand{\FTeV}{F_{\rm TeV}}
\newcommand{\LMaxTeV}{L_{\rm TeV,\, Max}}
\newcommand{\LMinTeV}{L_{\rm TeV,\, Min}}
\newcommand{\EpsInfTeV}{\epsilon_{\rm TeV}^{\rm inf}}
\newcommand{\EpsSupTeV}{\epsilon_{\rm TeV}^{\rm sup}}
\newcommand{\LMWTeV}{L^{\rm MW}_{\rm TeV}}
\newcommand{\PhiTotTeV}{\Phi^{\rm tot}_{\rm TeV}}
\newcommand{\PhiTeV}{\Phi_{\rm TeV}}

\newcommand{\ETeV}{E_{\rm TeV}}
\newcommand{\betaTeV}{\beta_{\rm TeV}}

\newcommand{\LGeV}{L_{\rm GeV}}

\newcommand{\EpsInfGeV}{\epsilon_{\rm GeV}^{\rm inf}}
\newcommand{\EpsSupGeV}{\epsilon_{\rm GeV}^{\rm sup}}

\newcommand{\PhiTotGeV}{\Phi^{\rm tot}_{\rm GeV}}
\newcommand{\PhiGeV}{\Phi_{\rm GeV}}
\newcommand{\PhiGeVTh}{\Phi_{\rm GeV}^{\rm th}}
\newcommand{\PhiGeVUNRES}{\Phi_{\rm GeV}^{\rm NR}}

\newcommand{\betaGeV}{\beta_{\rm GeV}}

\newcommand{\LX}{L_{\rm X}}
\newcommand{\FX}{F_{\rm X}}
\newcommand{\PhiX}{\Phi_{\rm X}}
\newcommand{\EX}{E_{\rm X}}

\newcommand{\Rphi}{R_{\Phi}}

\newcommand{\phiUnit}{\,{\rm cm}^{-2}\,{\rm s}^{-1}}
\newcommand{\lUnit}{\,{\rm erg}\,{\rm s}^{-1}}

\shorttitle{}
\shortauthors{}

\begin{document}

\title{The contribution of Galactic TeV pulsar wind nebulae to Fermi Large Area Telescope diffuse emission}

\correspondingauthor{Vittoria Vecchiotti}
\email{vittoria.vecchiotti@gssi.it}

\author{Vittoria Vecchiotti}
\affiliation{Gran Sasso Science Institute, 67100 L'Aquila, Italy}
\affiliation{INFN, Laboratori Nazionali del Gran Sasso, 67100 Assergi (AQ),  Italy}

\author{Giulia Pagliaroli}
\affiliation{Gran Sasso Science Institute, 67100 L'Aquila, Italy}
\affiliation{INFN, Laboratori Nazionali del Gran Sasso, 67100 Assergi (AQ),  Italy}

\author{Francesco Lorenzo Villante}
\affiliation{INFN, Laboratori Nazionali del Gran Sasso, 67100 Assergi (AQ),  Italy}
\affiliation{University of L'Aquila, Physics and Chemistry Department, 67100 L'Aquila, Italy}

\begin{abstract}
The large-scale diffuse $\gamma-$ray flux observed by Fermi Large Area Telescope (Fermi-LAT) in the 1-100 GeV energy range, parameterized as $\propto E^{-\Gamma}$, has a spectral index $\Gamma$ that depends on the distance from the Galactic center. 
This feature, if attributed to the diffuse emission produced by cosmic rays interactions with the interstellar gas, can be interpreted as the evidence of a progressive cosmic ray spectral hardening towards the Galactic center.
This interpretation challenges the paradigm of uniform cosmic rays diffusion throughout the Galaxy.
We report on the implications of TeV Pulsar Wind Nebulae observed by the High Energy Stereoscopic System (H.E.S.S.) Galactic Plane Survey in the 1-100 TeV energy range for the interpretation of Fermi-LAT data.
We argue that a relevant fraction of this population cannot be resolved by Fermi-LAT in the GeV domain providing a relevant contribution to the large-scale diffuse emission, ranging within $\sim 4\%-40\%$ of the total diffuse $\gamma$-ray emission in the inner Galaxy.
This additional component may account for a large part of the spectral index variation observed by Fermi-LAT, weakening the evidence of cosmic ray spectral hardening in the inner Galaxy. 

\end{abstract}


\section{Introduction} \label{sec:outline}

Cosmic Rays (CRs) with energy below $\sim1$ PeV are believed to originate in the Milky Way and to spread in the entire Galaxy due to diffusion in local magnetic fields~\citep{Gabici:2019jvz}.
The diffuse $\gamma$-ray emission, produced by interaction of CRs with the gas contained in the galactic disk, carries information on the energy distribution of CRs in different regions of the Galaxy.

Recent observations at GeV energies performed by Fermi-LAT suggest that the  hadronic diffuse gamma-ray emission, parameterized as $\propto E^{-\Gamma}$, has a spectral index $\Gamma$ in the inner Galaxy which is smaller by an amount $\sim -0.2$ than the value observed at the Sun position \cite{Pothast:2018bvh}. 
This feature can be considered as the indirect evidence of a progressive CR spectral hardening towards the Galactic center \cite{Yang:2016jda,Acero:2016qlg}.%
This conclusion, however, challenges standard implementations of the CR diffusion paradigm, in which uniform diffusion throughout the Galaxy is assumed, and would require a more complex description of CR transport \citep{Recchia:2016bnd, Cerri:2017joy}.
It is thus extremely important to consider any possible alternative explanations of Fermi-LAT results \cite{Nava:2017qun}.

An essential step for the observational identification of CR diffuse emission, is the evaluation of the cumulative flux produced by sources which are too faint to be resolved by Fermi-LAT.
These sources are not individually detected but give rise to a large scale diffuse flux superimposed to that produced by CR interactions.
%
%
%
To investigate the role of this additional component recent works \cite{Acero:2016qlg,Pothast:2018bvh} performed a source population study concluding that the diffuse flux associated to unresolved sources is not large enough to explain the spectral anomaly being below 3\% at 1 GeV (20\% at $\simeq 100$ GeV) of the total observed diffuse emission.
Both studies are tuned on the 3FGL catalog. As a consequence, they reproduce the population of Galactic sources observed in the GeV energy domain which is largely dominated by Pulsars.
These objects have $\gamma-$ray spectra with exponential cutoff at few GeV and are expected to provide a negligible contribution to observed emission at $E\ge 10\;$GeV. 

%
In the last decade, Imaging Atmospheric Cherenkov Telescopes (IACT), like H.E.S.S. \citep{Aharonian:2005kn}, MAGIC \citep{Aleksic:2014lkm} and VERITAS \citep{Weekes:2001pd}, and air shower arrays, such as Argo-YBJ \citep{Bartoli:2013qxm}, Milagro \citep{Atkins:2004yb} and HAWC \citep{Abeysekara:2015qba,Abeysekara:2017hyn,PhysRevLett.124.021102}, provided a detailed description of Galactic $\gamma-$ray emission in the energy range $0.1-100\,{\rm TeV}$.
%
%
The emerging picture is that TeV Galactic sky is dominated  by a population  of  bright  sources  powered by pulsar activity, such as pulsar wind nebulae (PWNe) \citep{Abdalla:2017vci} or TeV halos \citep{Linden:2017blp,Sudoh:2019lav,Giacinti:2019nbu}, whose properties can be effectively constrained by observations at TeV energies \citep{Cataldo:2020qla,Steppa:2020qwe}.
These  objects  are  clearly expected  to  emit  also  in  the  GeV  energy  domain where, however, population studies are more difficult because different kinds of sources dominate the observed emission.

In this paper, we took advantage of the constraints provided by H.E.S.S. Galactic Plane Survey (HGPS) to discuss the implications of TeV PWNe for the interpretation of Fermi-LAT data in the GeV domain.
We quantify the contribution of unresolved TeV PWNe to large scale diffuse emission observed by Fermi-LAT at different distances from the Galactic center.
%
We show that the inclusion of this  additional  component can strongly affect the reconstructed CR energy distribution from Fermi-LAT data,
weakening the evidence of a progressive hardening of the cosmic-ray spectrum toward the Galactic center.
\\

\section{Results and Discussion}
\label{sec:Results}

Pulsar wind nebulae are expected to contribute to $\gamma$ observations both in the GeV and TeV energy domains. 
We indicate with $\PhiGeV$ ($\PhiTeV$) the integrated source flux in the energy range $1-100\,{\rm GeV}$ ($1-100\,{\rm TeV}$) probed by Fermi-LAT (H.E.S.S.).
We assume that all the sources in the considered population have approximately the same emission spectrum, described by a broken power-law with different spectral indexes $\betaGeV$ and $\betaTeV$ in the GeV and TeV energy domain and with a transition energy  $E_0 = [0.1-1.0]\,{\rm TeV}$ located between the ranges probed by Fermi-LAT and H.E.S.S.. 
%
%
%
At high energies ($E\ge E_0$), we allow the source spectral index to move inside the range $\betaTeV = [1.9 - 2.5]$ measured by H.E.S.S. \citep{H.E.S.S.:2018zkf} for identified PWNe, see Appendix A. 
The index $\betaGeV$ is instead determined by requiring realistic values for the parameter $\Rphi$, defined as the ratio
%
\begin{equation}
\label{eq:Rphi}
\Rphi \equiv \frac{\PhiGeV}{\PhiTeV}
\end{equation}
between fluxes emitted by a given source in different energy domains.
As it is discussed in Sect.~\nameref{sec:method}, we obtain a consistent description of the HGPS and the Fermi Large Area Telescope Fourth Source Catalog data released two (4FGL-DR2) for $\Rphi=\left[250-1500\right]$ that corresponds  to $\betaGeV$ inside the global range $\betaGeV=[1.06-2.19]$ (see Eq.~\ref{Rphi}  for the general relationship between the spectral parameters in our analysis).
The assumed source spectrum
can be further validated by considering the average observational properties of PWNe observed by Fermi-LAT and H.E.S.S. in the GeV/TeV domain, see Sect \nameref{sec:method} for details. Moreover, the corresponding spectral shapes are consistent with theoretical predictions for $\gamma$-ray emission from PWNe \cite{Bucciantini:2010pd,Torres:2014iua}.

\begin{figure*}[!ht]
\begin{center}
\includegraphics[width=7cm]{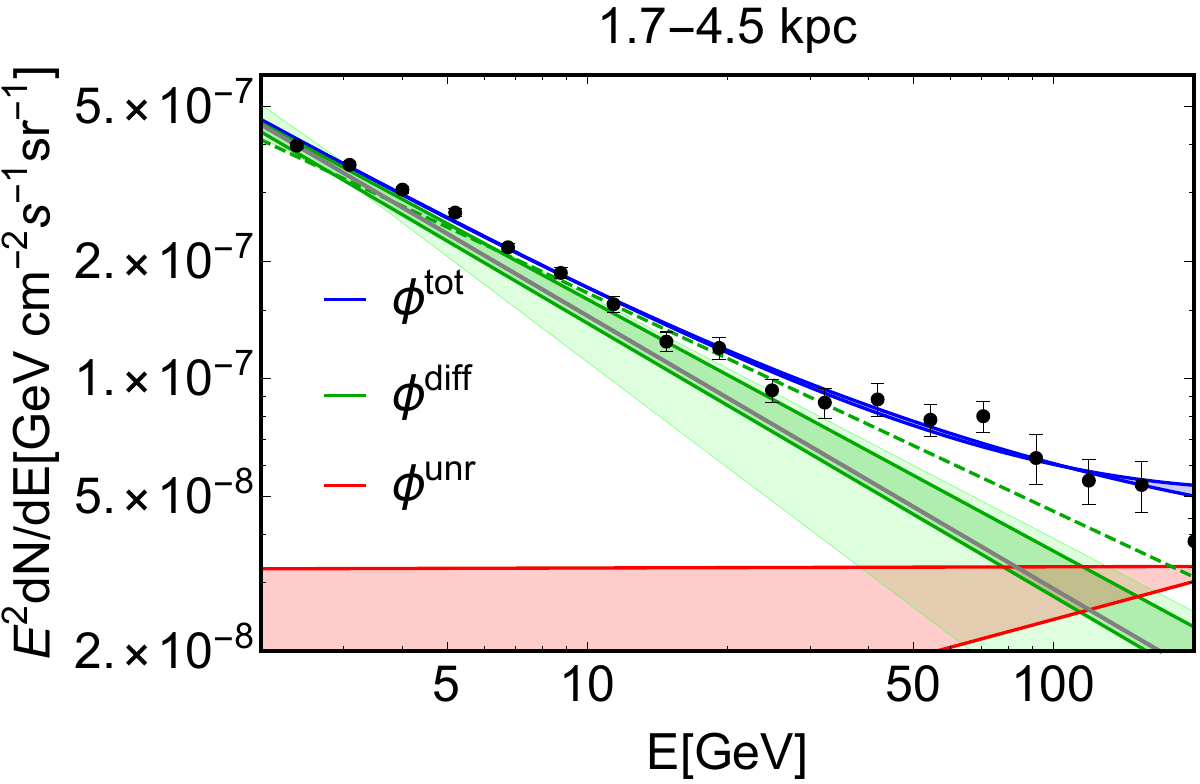}
\includegraphics[width=7cm]{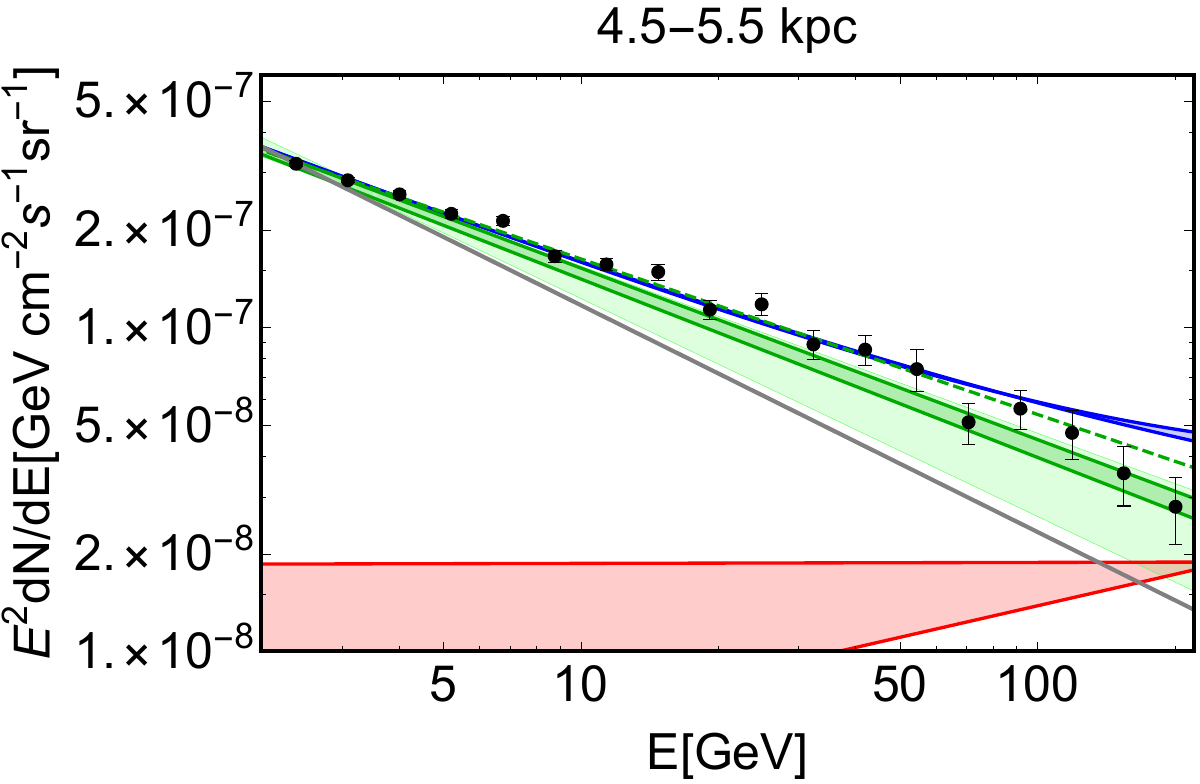}\\
\includegraphics[width=7cm]{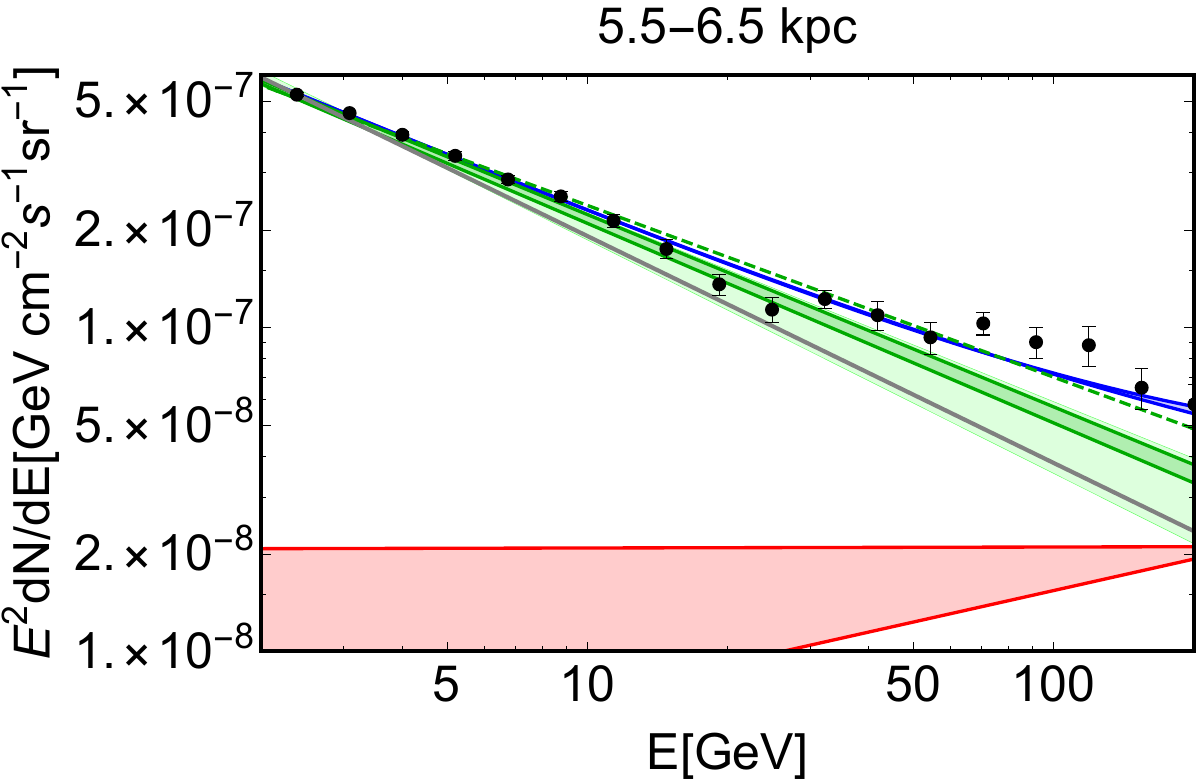}
\includegraphics[width=7cm]{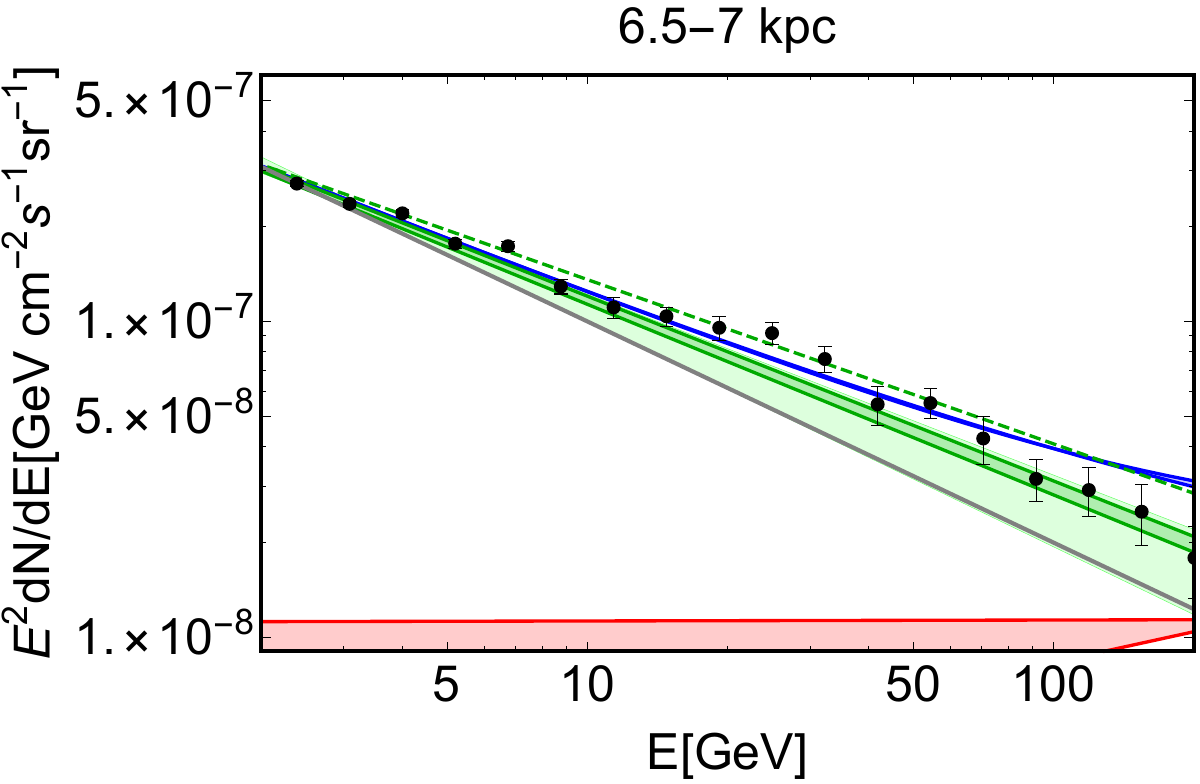}\\
\includegraphics[width=7cm]{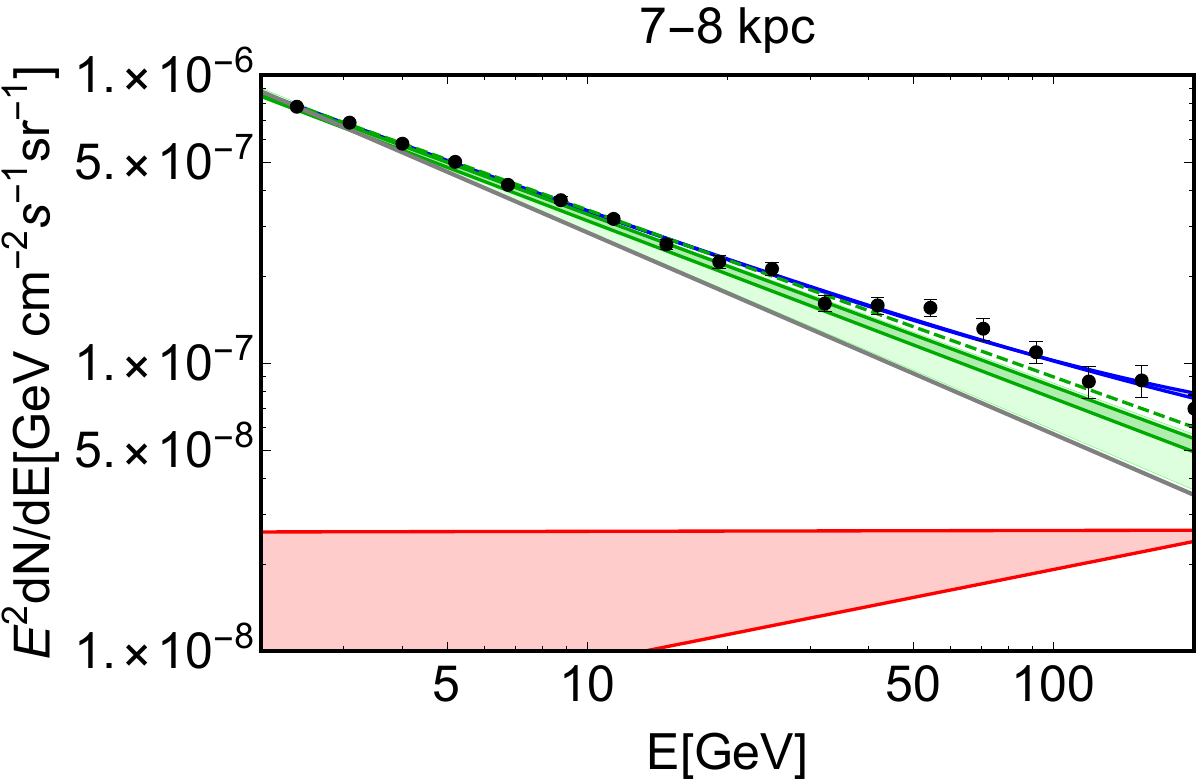}
\includegraphics[width=7cm]{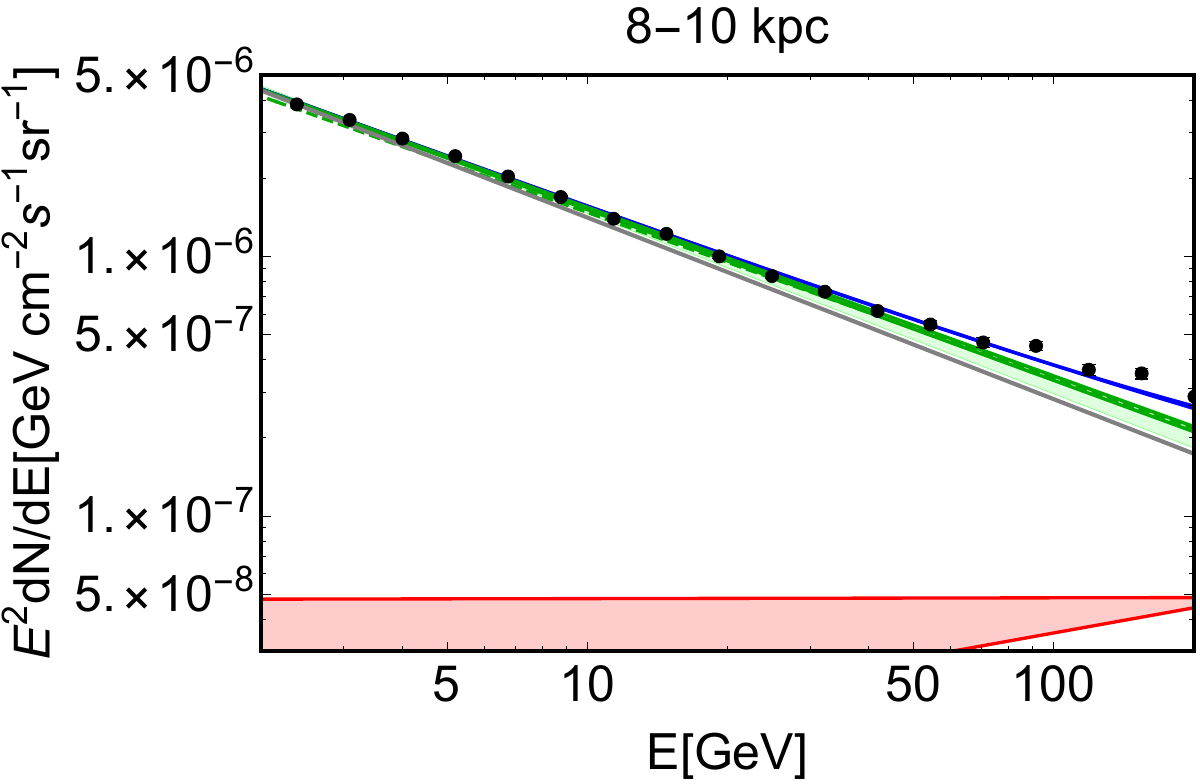}\\
\includegraphics[width=7cm]{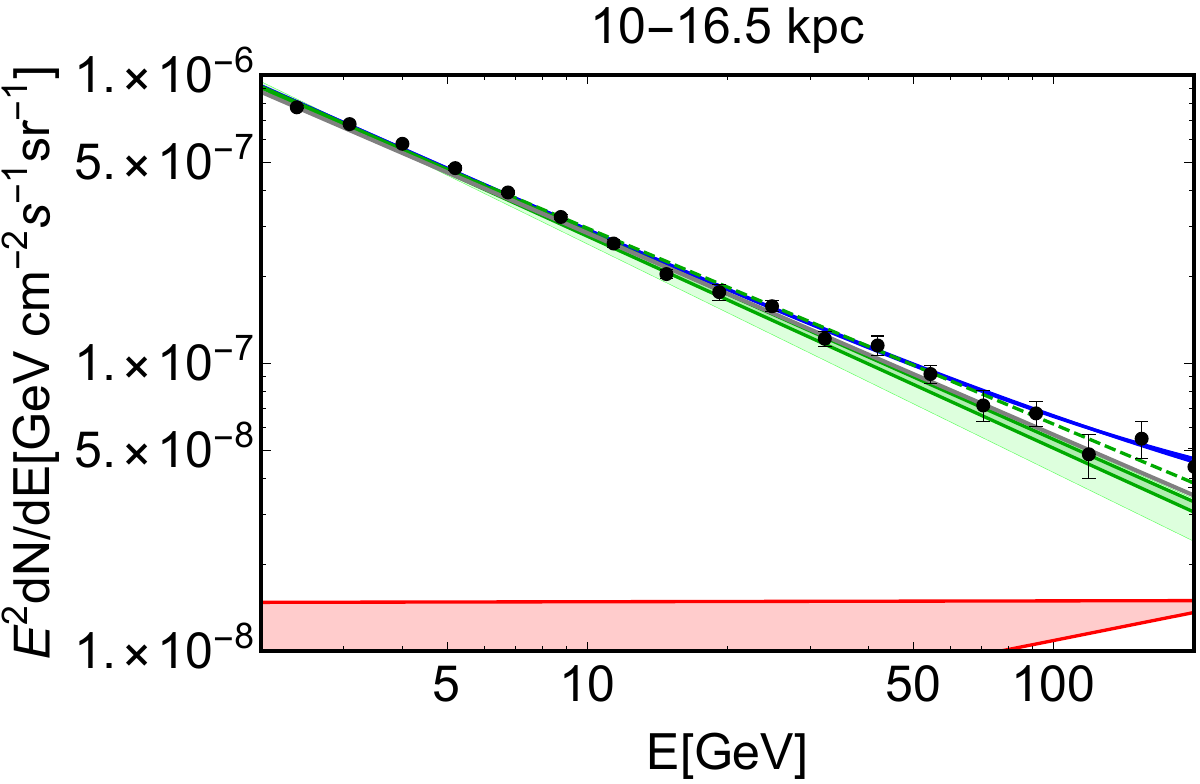}
\includegraphics[width=7cm]{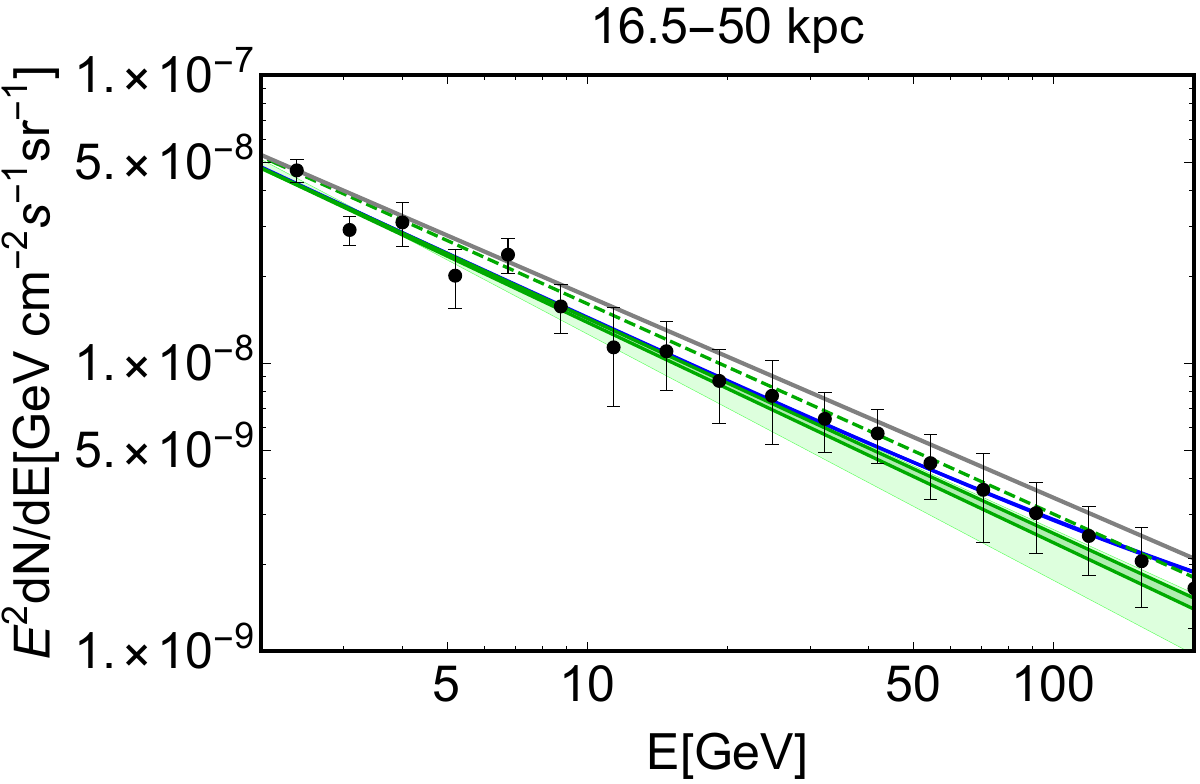}
\caption{\small\em 
{\bf The diffuse gamma-ray emission as a function of energy in different galactocentric rings.}
Black data points show the total diffuse $\gamma$-ray emission associated with interstellar gas measured by Fermi-LAT in each galactocentric ring \citep{Pothast:2018bvh}, the error bars represent the statistical error. 
The red bands represent the predicted contribution of unresolved TeV pulsar wind neubulae for $\alpha=1.8$, $E_0=0.8$ TeV, $\betaTeV=2.4$ and $\Rphi$ ranging between $250-1500$. 
Green lines show the diffuse cosmic ray emission inferred by fitting the data with (solid) and without (dashed) including the pulsar wind nebulae contribution. 
The dark green bands show the systematic error produced by variations of the flux ratio $\Rphi$.
Light green bands show the total systematical uncertainty obtained when $E_{0}$ and  $\betaTeV$ are also allowed to vary.
Blue lines represent the total gamma fluxes predicted as a function of the energy for $\alpha=1.8$, 
$E_0=0.8$ TeV, $\betaTeV=2.4$ and $\Rphi$ ranging between $250-1500$. 
The gray lines show a power-law with an index of 2.7 for comparison.
%
%
\label{fig:comp_with_data_alpha18}}
\end{center}
\end{figure*} 

\subsection{The PWNe population in the TeV domain.}
The properties of the considered source population can be constrained by observation in the TeV energy domain.
Following a previous work \cite{Cataldo:2020qla}, PWNe distribution is described by:
\begin{equation}
\frac{dN}{d^3 r\,d\LTeV} = \rho\left({\bf r} \right) \YTeV \left(\LTeV\right)  
\label{SpaceLumDist}
\end{equation}
where $r$ indicates the source distance from the Galactic Center. 
The function $\rho({\bf r})$ describes the spatial distribution of the sources and it is conventionally normalized to one when integrated in the entire Galaxy. It is assumed to be proportional to the pulsar distribution in the Galactic plane \cite{Lorimer:2006qs}.
The source density along the direction perpendicular to the Galactic plane is assumed to scale as $\exp \left(-\left|z \right|/H\right)$ where $H=0.2\ {\rm kpc}$ represents the thickness of the Galactic disk. 

The function $\YTeV(\LTeV)$ gives the source intrinsic luminosity distribution in the TeV energy domain. 
%
%
It is parameterized as a power-law:
\begin{equation}
\YTeV(\LTeV)=\frac{{R \, \tau \, (\alpha-1)}}{\LMaxTeV}\left(\frac{\LTeV}{\LMaxTeV}\right)^{-\alpha}
\label{LumDist1} 
\end{equation}
that extends in the luminosity range $\LMinTeV \le \LTeV \le
\LMaxTeV$ \cite{Strong:2006hf}.
%
This functional form, that is generically adopted in population studies, is naturally obtained for a population of {\em fading} sources, such as PWNe or TeV Halos, produced with a constant rate $R$ and having intrinsic luminosity that decreases over a time scale $\tau$, see Methods for details.
%

Previous analyses on the subject  \citep{Acero:2016qlg, Pothast:2018bvh} have been performed under the assumption that the index of the luminosity distribution is $\alpha=1.8$ because this leads to a good description of observational data.
We conform to this choice for our reference case and we note that the value $\alpha=1.8$ is also obtained for a population of pulsar-powered sources, if the efficiency of TeV emission is correlated to spin-down power as suggested by H.E.S.S. data \cite{Abdalla:2017vci}. 
%
%
However, to show the dependence of our results on the performed assumptions, we also consider the alternative hypothesis $\alpha=1.5$ 
that is obtained by postulating that the efficiency of TeV emission is constant in time.
%

By fitting the flux, latitude and longitude distribution of bright sources in the HGPS catalog (and assuming that the PWNe birth rate is equal to that of core-collapse SN explosions (this analysis is only sensitive to the  product $R\tau$. A smaller PWN formation rate can be balanced by a higher value of $\tau$ (and viceversa) with no consequences for the present discussion), i.e.  $R=0.019\;{\rm yr}^{-1}$), one obtains  $\LMaxTeV=6.8\times 10^{35}\lUnit$ ($\LMaxTeV = 4.9 \times 10^{35}\lUnit$) and $\tau = 0.5 \times 10^3\,{\rm y}$ ($\tau =1.8\times 10^3\,{\rm y}$) for $\alpha=1.8$ ($\alpha=1.5$) \citep{Cataldo:2020qla}.
The fit to HGPS sources permits us to constrain the cumulative flux $\PhiTotTeV$ produced by PWNe population in the TeV domain with $\sim 30\%$ statistical accuracy, being it equal to $\PhiTotTeV = \left(5.9^{+1.8}_{-1.5}\right)\times 10^{-10} \phiUnit$ for $\alpha=1.8$.
This estimate does not critically depend on the adopted assumptions (e.g. the source space distribution,
the Galactic disk thickness, the source physical dimensions, etc.).
The largest effect is obtained by modifying the luminosity index $\alpha$. 
We get e.g. $\PhiTotTeV = \left(3.8^{+1.2}_{-1.1}\right)\times 10^{-10} \phiUnit$ for $\alpha = 1.5$ that corresponds to $\sim 35\%$ reduction with respect to the reference case $\alpha =1.8$ \cite{Cataldo:2020qla}.

 The above results have been obtained by assuming $\betaTeV=2.3$ which corresponds to the average spectral index of sources observed by H.E.S.S.. 
%
%
 It should be remarked, however, that the adopted value of $\betaTeV$ has no effects on the cumulative flux $\PhiTotTeV$, neither on the distribution $dN/d\PhiTeV$ of sources as function of flux in the TeV domain. These quantities are thus directly determined by observational data, independently on assumptions for the source emission spectrum.

%
%
%

\subsection{The total and unresolved emission in the  GeV domain.}
The total flux produced at Earth by TeV PWNe population in the GeV domain depends on the parameter $\Rphi$ and it is given by:
\begin{equation}
\PhiTotGeV = \Rphi\, \PhiTotTeV.
\label{Eq:PhiTotGeV}
\end{equation}
 The parameter $\Rphi$ also determines the distribution of sources as a function of the flux they emit at GeV energies, according to:
\begin{equation}
\frac{dN}{d\PhiGeV} = \frac{1}{\Rphi} \frac{dN}{d\PhiTeV}\left(\PhiGeV/\Rphi  \right) ~.
\label{Eq:dNdphiGeV}
\end{equation}
Faint sources cannot
be individually resolved by Fermi-LAT and contribute to the large scale diffuse emission observed by this experiment. 
The unresolved contribution can be calculated as:
\begin{equation}
\PhiGeVUNRES = \int_{0}^{\PhiGeVTh} d\PhiGeV \; \PhiGeV \,  \frac{dN}{d\PhiGeV}  
\label{Eq:PhiGeVUnr}
\end{equation}
where $\PhiGeVTh$ is the Fermi-LAT detection threshold. 
%
For objects contained
in the Galactic plane, this is estimated as $\PhiGeVTh = 10^{-9} \phiUnit$ \cite{Acero:2015gva}   by looking at the turnover of the observed source number as a function of the photon flux above $1\;{\rm GeV}$ (see their Fig.~24, panel (a)).
By considering that the flux distribution scales as $dN/d\PhiTeV \propto \PhiTeV^{-\alpha}$ for $\PhiTeV\to 0$ (see Sect.~\nameref{sec:method}), we expect that $\PhiGeVUNRES\propto \Rphi^{\alpha-1}$.
%
%
We remark that the total and unresolved fluxes, $\PhiTotGeV$ and $\PhiGeVUNRES$, only depend on $\Rphi$ and are independent on the assumed values for the spectral parameters $E_0$ and $\betaTeV$.

In the last line of Tab.~\ref{Tab1}, we give the flux $\PhiGeVUNRES$ produced by PWNe that are not resolved by Fermi-LAT 
for the two extreme values $\Rphi=250$ and $1500$.
These fluxes are compared with the large scale diffuse emission associated with interstellar gas $\Phi^{\rm diff}_{\rm GeV}$ detected by Fermi-LAT (see second column in Tab.\ref{Tab1}) in the $1-100$~GeV energy range and determined
in Pothast {\em et al.} \cite{Pothast:2018bvh} by using 9.3 years of Fermi-LAT Pass 8 data. The energy integrated fluxes have been obtained by interpolating the experimental points  and integrating in the energy range $1-100$ GeV.
We see that unresolved emission by PWNe corresponds to a fraction $\sim 3\%$ (for $\Rphi=250$) and $\sim 11\%$ (for $\Rphi=1500$) of the diffuse gamma-ray emission associated with interstellar gas.
The above results are obtained by assuming that the source luminosity distribution index is $\alpha=1.8$ to conform with previous analyses on the subject \citep{Acero:2016qlg,Pothast:2018bvh} that have been performed under this hypothesis. 
Results for $\alpha=1.5$ are smaller and are reported in the last two columns of Tab.~\ref{Tab1}.

In order to probe the radial dependence of the PWNe contribution, we repeat our calculations by considering the Galactocentric rings adopted in Pothast {\em et al.} \cite{Pothast:2018bvh}. 
The flux produced by unresolved TeV PWNe in each ring is compared with the Fermi-LAT diffuse emission from the same region.
As we see from Tab.~\ref{Tab1}, the unresolved contribution becomes more relevant in the central rings, due the fact that the source density (and the average distance from the Sun position) is larger.
In the most internal region ($1.7\le r \le 4.5\, {\rm kpc}$), unresolved sources account for $\sim 9\%$ ($\sim 36\%$) of the Fermi-LAT diffuse emission associated with interstellar gas for $\Rphi=250$ ($\Rphi=1500$) and $\alpha=1.8$. We do not consider the central region $r \le 1.7$ because it is affected by large systematic errors \cite{Pothast:2018bvh}.
This clearly shows that this component is not negligible and cannot be ignored in the interpretation of Fermi-LAT diffuse emission data.
%

\subsection{Spectral analysis}
\begin{figure}[htbp]
\begin{center}
\includegraphics[width=0.45\textwidth]{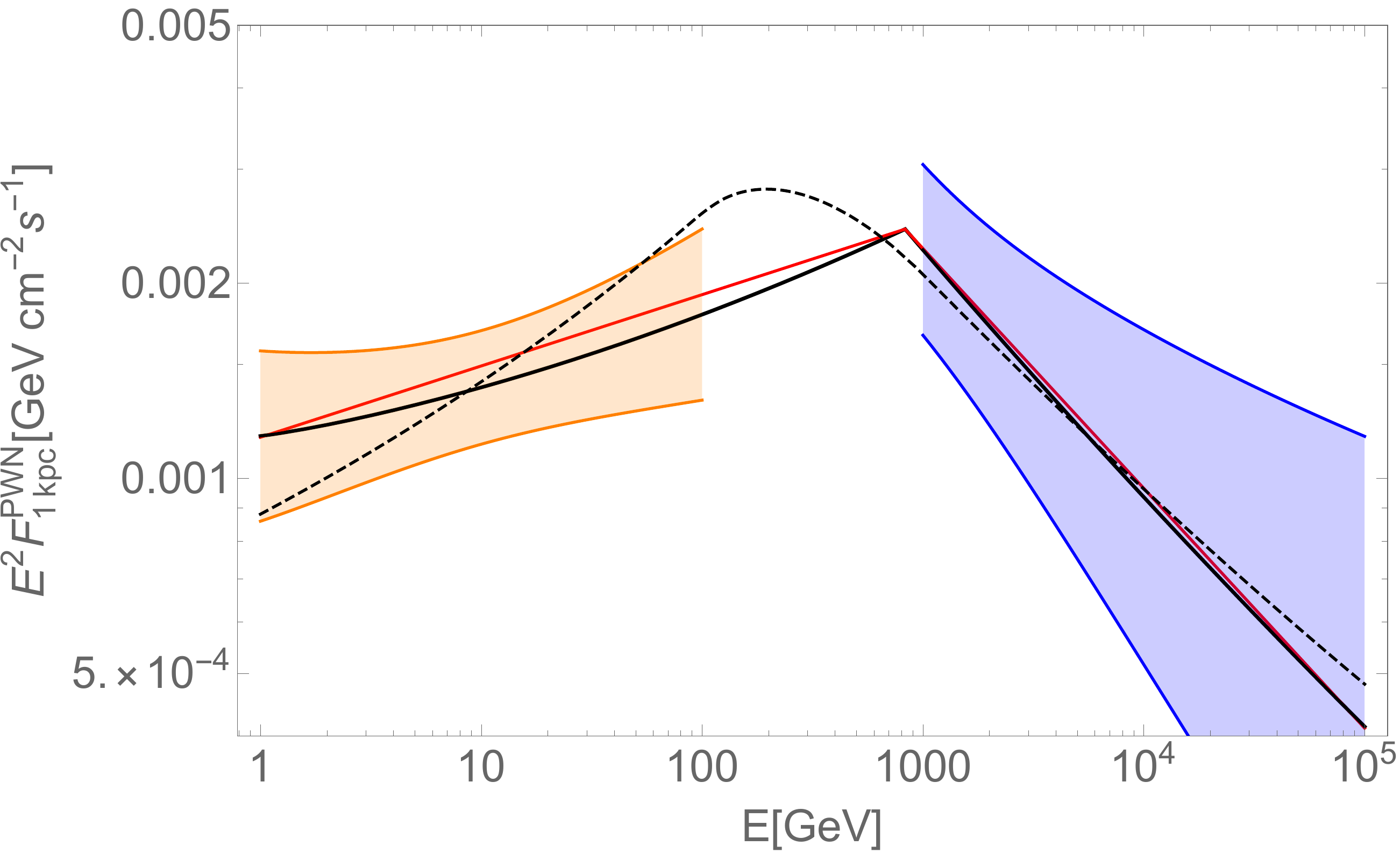}\\
\caption{\small\em {\bf Pulsar wind nebulae cumulative spectrum.} The cumulative spectral energy distribution of the PWNe observed both by Fermi-LAT and H.E.S.S. (black thick line). We show with a red line a broken power-law spectrum with an energy break  $E_{0} = 0.8$~TeV and spectral indexes $\beta_{\rm TeV} = 2.4$ and $\beta_{\rm GeV} = 1.89$.  The shaded bands are obtained by propagating statistical and systematic uncertainties in the source spectra given in 4FGL-DR2 \citep{Fermi-LAT:2019yla} (orange band) and HGPS \citep{H.E.S.S.:2018zkf} (blue band) catalogs. The dashed black line represents the average theoretical spectrum obtained by integrating over the spectral parameter space (see Sect. \nameref{susec:results}). It corresponds to the central values of $\Gamma_{BF}$ given in Tab.\ref{tab:gamma18} and it is normalized in order to produce the same number of photons in the TeV energy domain as the observed cumulative PWNe spectrum.} 
\label{fig:PWN}
\end{center}
\end{figure}

\begin{figure}[t]
\includegraphics[width=0.45\textwidth]{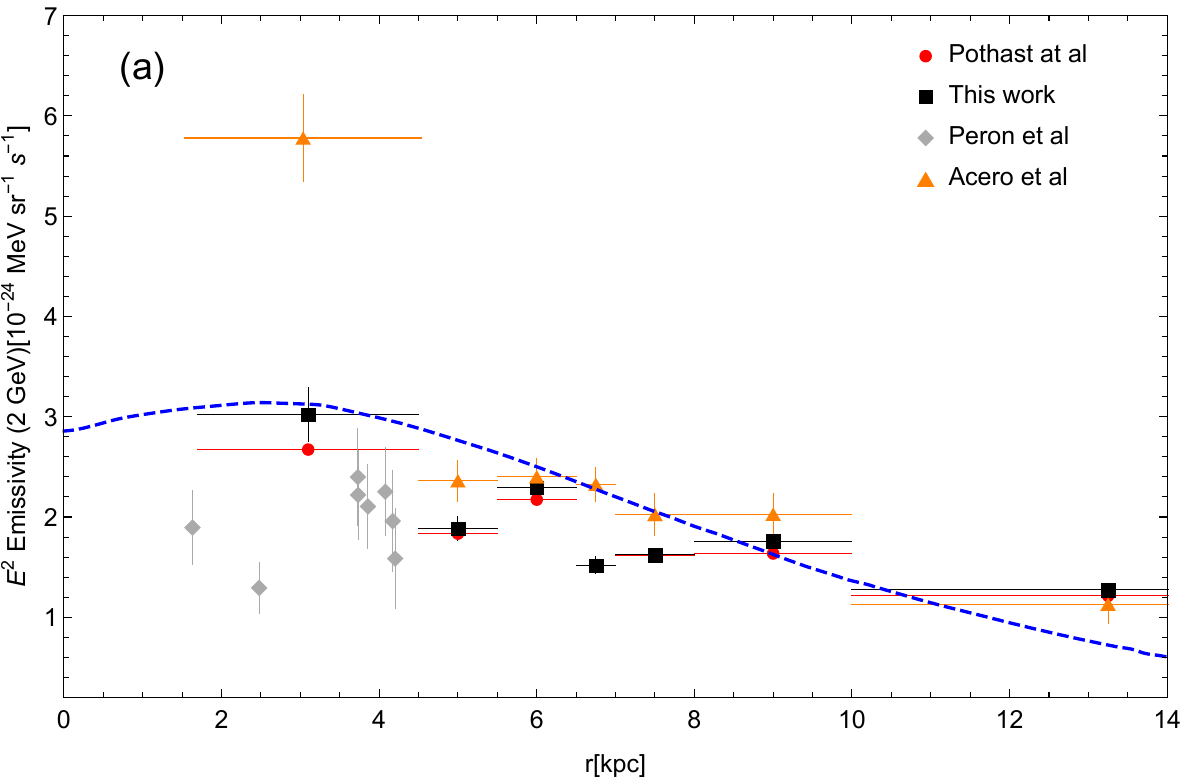}\\
\includegraphics[width=0.45\textwidth]{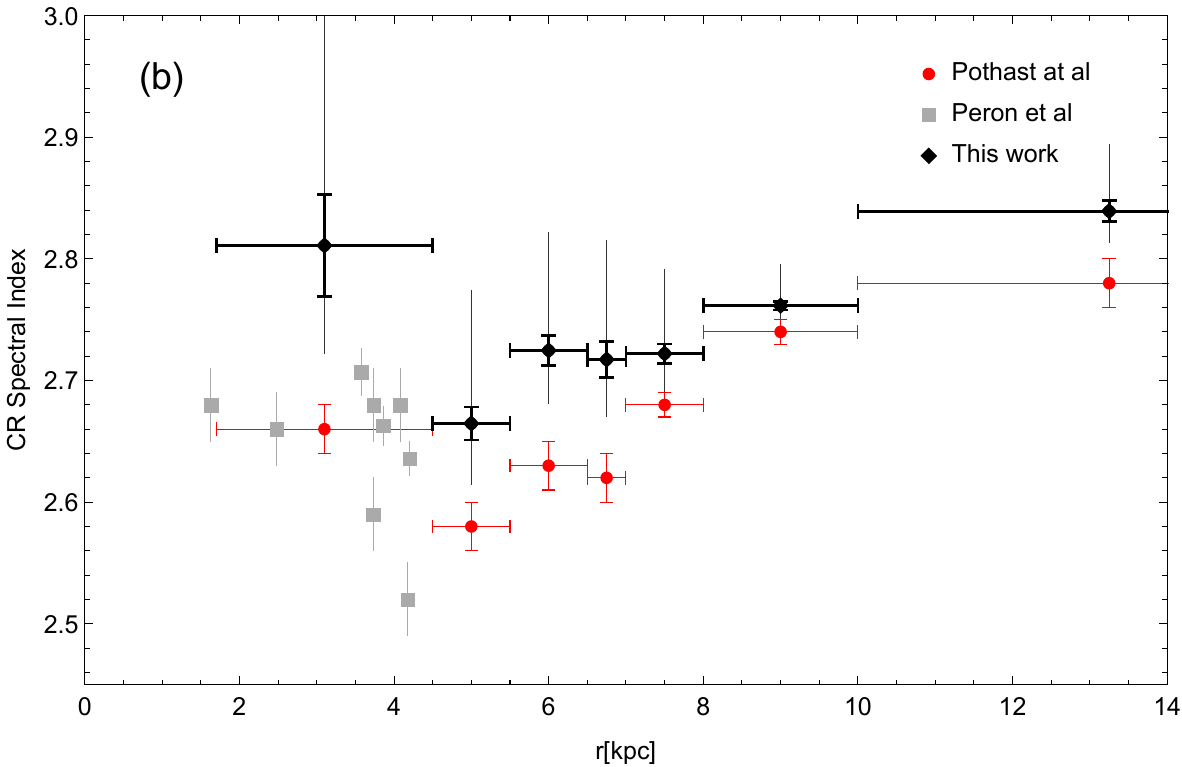}
\caption{\small\em {\bf Gamma-rays emissivity and cosmic ray proton spectral index in different galactocentric rings:} The gamma-ray emissivity (a) and the CR spectral index (b) obtained in this work (black points) compared with the ones in Peron {\em et al.} \cite{Peron:2021lwb} (gray points), Pothast {\em et al.} \cite{Pothast:2018bvh} and in  Acero {\em et al.}\cite{Acero:2016qlg} (orange points). 
The error bars for black points are obtained by summing in quadrature statistical and systematical uncertainties, see Tab.\ref{tab:gamma18}.  In particular, thin error bars show the systematic uncertainties conservatively estimated, while thick error bars only include statistical uncertainties.
The blue dashed line represents the CR distribution predicted by the GALPROP code \citep{Strong:2004td} normalized at $8.5\;kpc$ at the emissivity value obtained in this work in the ring $8-10\;kpc$.
\label{fig:IndexEmissivity}}
\end{figure}

The effect of the unresolved TeV PWNe population on the determination
of CR diffuse emission spectral index is displayed in
Fig.~\ref{fig:comp_with_data_alpha18}. 
%
The purpose of this figure is not to discuss comprehensively the effects of parameters variations in our calculation.
Rather, our goal is to illustrate 
our approach 
and to explain why, despite the extremely large range of variation of the $\Rphi$ parameter (determining the PWNe integrated flux in the GeV domain), one still gets a prediction for the spectral index of CR diffuse emission.
For this reason, we fix the spectral parameters to the values that better reproduce the cumulative spectral energy distribution of the PWNe observed both by Fermi-LAT and H.E.S.S. (i.e. $\betaTeV=2.4$ and $E_0 = 0.8\,{\rm TeV}$, see Fig.\ref{fig:PWN}) and we only vary the flux ratio in the range $\Rphi=250-1500$.
On the other hand, the final results of our analysis reported in Tab.~\ref{tab:gamma18} and displayed in Fig.~\ref{fig:IndexEmissivity}, also take into account effects of possible variations of $E_0$ and $\betaTeV$.
%
%

Black data points in Fig.~\ref{fig:comp_with_data_alpha18} represent the total $\gamma-$ray flux associated with interstellar gas observed by Fermi-LAT in each galactocentric ring  in $25$ log-spaced energy bins between $0.34-228.65\;{\rm GeV}$ and in the latitude window $|b| < 20.25^{\circ}$. 
These data have been previously  fitted in Pothast {\em et al.} \cite{Pothast:2018bvh} with a single power-law $\propto E^{-\Gamma_{1}}$,
%
obtaining the green dashed lines reported in Fig.\ref{fig:comp_with_data_alpha18}.
The decrease of the best-fit spectral indexes $\Gamma_{1}$ in the inner rings with respect to the locally observed value, see second column of Tab.\ref{tab:gamma18}, has been considered as the evidence of a progressive large-scale hardening of CRs spectrum toward the Galactic Center.
The same conclusion was obtained by previous analyses on the subject \citep{Acero:2016qlg,Yang:2016jda} performed by using a similar approach.
One can get a visual perception of the situation by comparing the green dashed lines with the grey solid lines in Fig.\ref{fig:comp_with_data_alpha18} that describe power laws with spectral index fixed at the local value, i.e. $\sim2.7$, suitably normalized to reproduce the observed flux at $2$~GeV.

The above conclusion is only valid if unresolved source contribution is negligible, so that the total observed emission can be identified with the "truly" diffuse component produced by CRs interaction with interstellar matter. 
This assumption is, however, not adequate in the inner Galaxy, as it is shown with red solid lines in Fig.\ref{fig:comp_with_data_alpha18} that give the unresolved PWNe contribution as a function of energy for the reference case $\alpha=1.8$ and the two extreme values $\Rphi=250$ and $1500$.
The red shaded area can be considered as the systematical uncertainty associated to the parameter $\Rphi$. The effects of possible variations of $E_0$ and $\betaTeV$ on the unresolved PWNe emission are shown in the Supplementary Material, see Fig.\ref{fig:SpettriPWNe_Irr}.
The relevant point to note in this figure is that the GeV source spectral index $\betaGeV$
and the flux ratio $\Rphi$ are correlated, as it is discussed in Sect. \nameref{sec:method} (see Eq.~\ref{Rphi}).
As a result of this, PWNe unresolved emission in the inner Galaxy is either relatively large or has an hard spectrum, providing a contribution at $E\sim 100\,{\rm  GeV}$ that is almost independent on $\Rphi$.
This is the natural consequence of the fact that the source emission above 1 TeV is observationally fixed by HGPS data.  
This important piece of information cannot be neglected and it is included in our work. 
%

%
%
%
%

%
%

If we take unresolved PWNe emission into account, the evidence for CR spectral hardening in the inner Galaxy may be considerably weakened, as it is shown by the green thick solid lines in Fig.~\ref{fig:comp_with_data_alpha18}
that represent the component of the total diffuse gamma-ray flux that can be ascribed to CR interactions.
%
This is still parameterized as a single power-law $\propto
E^{-\Gamma_{BF}}$ (the number of degrees of freedom in the fit is not changed) but the total flux, described by blue lines in Fig.~\ref{fig:comp_with_data_alpha18}, is obtained as the sum of CR diffuse emission plus the unresolved PWNe contribution.
%
%
The best fit spectral indexes $\Gamma_{BF}$ of the truly diffuse
emission obtained in each ring are are larger and closer to the value measured at the Sun position with respect to those obtained in previous analyses
\citep{Pothast:2018bvh, Acero:2016qlg, Yang:2016jda} that do not take unresolved sources into account, see Tab.~\ref{tab:gamma18}. 
 Our results are mildly dependent on the flux ratio $\Rphi$, as it can be understood by looking at the dark green bands in Fig.~\ref{fig:comp_with_data_alpha18} that show the effects of varying this parameter in the range $\Rphi = [250-1500]$.
The light green bands also take into account possible variations of the spectral parameters $E_0$ and $\betaTeV$ and provide a conservative estimate of the total systematical error for CR diffuse emission, as it discussed in the next paragraph.


\subsection{Spectral index}
\label{susec:results}
In order to 
estimate the uncertainties in our approach, we repeat 
%
our analysis for different combinations of the spectral parameters $(R_{\Phi}, \, E_{0} , \, \beta_{\rm TeV})$.
The final results of our analysis are 
given in Tab.~\ref{tab:gamma18}. Here, the first errors describe systematic uncertainties, evaluated as the maximal variations of $\Gamma_{BF}$
that are obtained when $(R_{\Phi}, \, E_{0} , \, \beta_{\rm TeV})$ are
simultaneously varied in the 3-dim parameter space defined by the ranges 
$R_\Phi= \left[250-1500\right]$,  $E_0 = \left[0.1-1.0 \right] \, {\rm TeV} $  and $\beta_{\rm TeV}=\left[1.9 -2.5 \right]$.
We see that our conclusions are stable and not challenged by possible systematic effects connected with the assumed source spectrum.
It is important to remark that our estimates for systematic uncertainties are very conservative. 
The smaller (larger) values for $\Gamma_{BF}$ are e.g. obtained by assuming that {\em all} sources in the considered population have $\betaTeV=1.9$ ($\betaTeV=2.5$), $E_0 = 1.0\,{\rm TeV}$ ($E_0 = 0.1\,{\rm TeV}$) with a marginal dependence on the assumed $\Rphi$, i.e. they correspond to a physical situation that is extremely unlikely.
TeV PWNe in our Galaxy are indeed expected to have a distribution of spectral properties with compensating effects among extreme assumptions.
The central values for $\Gamma_{BF}$ given in Tab.~\ref{tab:gamma18} %
are obtained by integrating over the whole parameters space. We assume logarithmic uniform distributions for the spectral break position
and for the flux ratio,
while for $\betaTeV$ we consider a Gaussian distribution centered in $\betaTeV=2.4$ and with dispersion $0.15$ as reported in the HGPS catalog~\citep{H.E.S.S.:2018zkf}.%

%

%

In addition to the reference case $\alpha=1.8$ (third column) that is obtained by using the luminosity distribution index considered by previous analyses \citep{Acero:2016qlg,Pothast:2018bvh}, we also display the results obtained by assuming the alternative value $\alpha=1.5$ (last column). 
In this case, one obtains smaller effects on $\Gamma_{BF}$, coherently with the fact that unresolved PWNe emission in the GeV domain is smaller, see Tab.~\ref{Tab1}. 
It would be important to have further phenomenological and/or theoretical constraints on the $\alpha$ parameter for future analyses.

The results of our reference case ($\alpha=1.8$) are compared with those given by other analyses in Fig.~\ref{fig:IndexEmissivity} where we show the 
$\gamma$-ray emissivity  
per H atom at 2 GeV  (a) which is a proxy of the CR spatial distribution in
the Galaxy, and the CR proton spectral index (b), obtained by adding
0.1 to the spectral indexes of the truly-diffuse gamma emission
\citep{Kelner:2006tc}.
Black points show the results of our work that are compared to those
given by Pothast {\em et al.}  \cite{Pothast:2018bvh} (red points) and Acero {\em et al.}  \cite{Acero:2016qlg}
(orange points). 
We also show with grey points the results obtained by
Peron {\em et al.} \cite{Peron:2021lwb} by studying $\gamma-$ray emission in the direction of
giant molecular clouds. 
The thin error bars (for the black points) show the systematic uncertainties conservatively estimated as discussed above
 while the thick error bars only include statistical uncertainties.
We see that the emissivity calculated in this work is in good agreement with that obtained by
Pothast {\em et al.} \cite{Pothast:2018bvh}  and Peron {\em et al.}  \cite{Peron:2021lwb}.
This is not surprising because we don't expect any significant effect at $2$ GeV due to the presence of unresolved sources.  
The three data-sets agree quite well with theoretical
expectations for the CR distribution from GALPROP code \citep{Strong:2004td} (dashed blue
line. The theoretical CR distribution is shown e.g. in Fig.~8 of \cite{Acero:2016qlg} where the specific GALPROP configuration is also given.
It basically coincides with the solution of 3D isotropic diffusion equation with uniform diffusion
coefficient, stationary CR injection and infinite smearing radius, as it is e.g. shown in Fig.~1 of \cite{Cataldo:2019qnz}.).
The inclusion of unresolved PWNe strongly affects the CR spectral index that can be increased up to $0.18$ in the central ring adjusting it to the locally observed value, i.e. $\sim 2.8$.
The cosmic ray reconstructed spectrum still shows a residual difference with the local value in the other rings.
We see, however, that unresolved PWNe naturally account for a
large part of the spectral index variation as a function of $r$ that has been reported by previous analyses, weakening considerably the evidence for CR spectral hardening in the inner Galaxy.

\subsection{Conclusions}

The TeV Galactic sky is dominated by a population of bright young PWNe whose properties are constrained by present H.E.S.S. Galactic Plane Survey (HGPS) data.
%
%
We predict the cumulative emission produced by this population in the GeV domain within a phenomenological model that is based on the average spectral properties of PWNe. This phenomenological description could be improved in the future adopting a more refined PWN spectrum as the one in Fiori {\em et al.} \citep{Fiori:2022nhu} appeared during the reviewing procedure of this work. 
We argue that a relevant fraction of the TeV PWNe population cannot be resolved by Fermi-LAT.
The $\gamma$-ray flux due to unresolved TeV PWNe and the truly diffuse emission, due to CR interactions with the interstellar gas, add up contributing to shape the radial and spectral behaviour of the total diffuse $\gamma$-ray emission observed by Fermi-LAT.
The spatial distribution of TeV PWNe, peaking around $r=4$ kpc from the Galactic Center, combined with the detector flux threshold modulate the relative contribution of unresolved sources in different Galactocentric rings. 
In particular the relevance of this component increases in the inner rings where the total diffuse emission has a different spectral distribution with respect to the local one. 
Previous analyses neglected the contribution due to unresolved PWNe and  interpreted the observed spectral behaviour of the total diffuse emission as an indirect evidence for CR spectral hardening toward the Galactic center \citep{Acero:2016qlg,Yang:2016jda, Pothast:2018bvh}. 
We have shown that the emergence of PWNe unresolved component in the central region, which is characterized by an average spectral index $\betaGeV<2.$, can strongly affect this conclusion, by naturally accounting for (a large part of) the spectral index observed variation as a function of $r$. 
Our results could also solve the tension, discussed in Cataldo {\em et al.} \cite{Cataldo:2019qnz}, between total $\gamma$-ray emission measured by H.E.S.S., Milagro, Argo and HAWC and that obtained by implementing CR spectral hardening.

\section{Method}
\label{sec:method}

\paragraph{Flux and luminosity ratios}
The sources considered in this work are expected to contribute to observations both in the GeV and TeV energy domains. 
We indicate with $\PhiGeV$ ($\PhiTeV$) and $\LGeV$ ($\LTeV$) the integrated source flux and luminosity in the energy range $1-100\,{\rm GeV}$ ($1-100\,{\rm TeV}$) probed by Fermi-LAT (H.E.S.S.).
We assume for simplicity that all the sources in the considered population have approximately the same emission spectrum. This automatically implies that the ratio $\Rphi \equiv \PhiGeV/\PhiTeV$
between fluxes emitted in different energy domains by a given source is fixed.
The relationship between intrinsic luminosity and flux produced at
Earth is generically written as:
\begin{equation} 
    \PhiX=\frac{\LX}{4 \pi r^2 \EX}
\label{Phi}
\end{equation}
where $r$ is the source distance, $\EX$ is the average energy of emitted photons and ${\rm X\, = GeV,\, TeV}$ indicates the considered energy range. 
It is also useful to define the integrated emissivity $\FX \equiv \LX / \EX$ that corresponds to the total amount of photons emitted per unit time by a given source in the ${\rm X\, = GeV,\, TeV}$ energy domain.

\subsection{Source spectrum}
The source emission spectrum $\varphi(E)$ can have a different  behaviour at GeV and TeV energies.
We take this into account by parameterizing it with a broken power-law with different spectral indexes $\betaGeV$ and $\betaTeV$ in the GeV and TeV energy domain and with a transition energy  $E_0$ located between the ranges probed by Fermi-LAT and H.E.S.S..
Even if our approach is completely phenomenological, the postulated spectral  behaviour is expected from a theoretical point of view.  
We are indeed considering the hypothesis that  most  of  the  bright TeV  sources  are  young  PWNe  and/or  TeV halos \cite{Sudoh:2019lav}. 
%
In this scenario, the observed gamma-ray emission is produced by IC scattering of HE electron and positrons on background photons (CMB, starlight, infrared). In the Thompson regime, this naturally produces hard gamma-ray emission with spectral index $ \beta \sim (p+1)/2$ where $p$ is the electron/positron spectral index.
At TeV energy, it produces instead a softer gamma-ray spectrum either due to the Klein-Nishina regime $\beta \sim (p+1)$ or to electron/positron energy losses \citep{Bucciantini:2010pd,Torres:2014iua,Sudoh:2021avj}.
%
In the assumption of a broken power-law for the gamma ray spectrum, the flux ratio $\Rphi$, the energy break $E_0$ and the two spectral indexes $\betaGeV$ and $\betaTeV$ are not independent and are related by the following expression:
\begin{equation}
\label{Rphi}
\Rphi = \frac{1-\betaTeV}{1-\betaGeV}\;
\frac{\left[(\EpsSupGeV)^{1-\betaGeV}-(\EpsInfGeV)^{1-\betaGeV}\right]}
{\left[(\EpsSupTeV)^{1-\betaTeV}-(\EpsInfTeV)^{1-\betaTeV}\right]}
\end{equation}
where $\EpsInfGeV \equiv (1.0\,{\rm GeV}/E_0)$ and $\EpsSupGeV \equiv (100\,{\rm GeV}/E_0)$ ($\EpsInfTeV \equiv (1.0\,{\rm TeV}/E_0)$ and $\EpsSupTeV \equiv (100\,{\rm TeV}/E_0)$) are the lower and upper bounds of the GeV (TeV) energy domains.
The above relationship implements mathematically the fact that, if the source spectral behaviour at high energies is known (i.e. $\betaTeV$ and $E_0$ are fixed), then the flux ratio $R_\Phi$ is an increasing
function of $\betaGeV$. 
In other words, the harder is the spectrum at GeV energies, the smaller is
the integrated flux in the GeV domain.
In our analysis, we vary the parameters $(R_{\Phi}, \, E_{0}, \, \beta_{\rm TeV})$ in the 3-dim parameter space defined by the ranges 
$R_\Phi= \left[250-1500\right]$, 
$E_0 = \left[0.1-1.0 \right] \, {\rm TeV} $ 
and {\bf $\beta_{\rm TeV}=\left[1.9 -2.5 \right]$}.
The spectral index $\beta_{\rm GeV}$ of GeV emission is determined as a function of $(R_{\Phi}, \, E_{0}, \, \beta_{\rm TeV})$ by inverting Eq.~\ref{Rphi}
and it  globally spans the range   $\beta_{\rm GeV} =1.06-2.19$.
By repeating our analysis for different combinations $(R_{\Phi}, \, E_{0}, \, \beta_{\rm TeV})$, we determine the stability of our results against the assumed source spectrum and we estimate the systematic uncertainties produced by the scatter of spectral properties in the PWNe population.
%
%
%
%
%
%
%

The assumed source spectrum can be validated by considering the ensemble of PWNe firmly identified both in the 4FGL-DR2 and HGPS catalogs (12 objects, reported in Tab.~\ref{tab:pwn} of the Supplementary Material). 
%
Within this sample, the average values for $\Rphi$ and $\beta_{\rm GeV}$ are $1122$ and $1.89$, respectively. These values fall inside the ranges of variation for these parameters considered in our analysis.
The spectral break position $E_{0}$, estimated as the crossing point of the spectral fits given by Fermi-LAT and H.E.S.S. in the GeV and TeV domain respectively, falls inside the range $0.1$ TeV-few TeV for all the sources.
Finally, the characteristic spectral energy distribution (SED) of the PWNe population is estimated by calculating the cumulative spectrum of sources included both in 4FGL-DR2 and HGPS catalogs (black line in Fig.\ref{fig:PWN}).
This is obtained by considering the best-fit spectra of each source, as given by Fermi-LAT and H.E.S.S. for the respective energy ranges.
We have included sources with a known distance $D$ whose spectra have been weighted proportionally to their intrinsic luminosity (namely, they have been scaled by a factor $(D/1 {\rm kpc})^2$).
The shaded bands are obtained by propagating the statistical errors on the spectral parameters of each source and by assuming a 20\% (30\%) systematic uncertainty for the flux normalization and 0.1 (0.2) systematic uncertainty for the spectral indexes in 4FGL-DR2 (HGPS) catalog, respectively.   
The cumulative spectrum is well reproduced by a broken power-law with an energy break at $E_{0}\simeq 0.8$ TeV, spectral indexes $\beta_{\rm TeV} =  2.4$ and $\beta_{\rm GeV} = 1.89$, and $R_{\Phi}\simeq 770$ (red line in Fig.\ref{fig:PWN}).
This functional form is allowed within the parameter space considered in this paper.

%

\subsection{Luminosity distribution}
In the following, we focus on the TeV-luminosity function since this can be effectively constrained by HGPS observational results \citep{Cataldo:2020qla}. The function $\YTeV(\LTeV)$ is parameterized as described in Eq. \ref{LumDist1}.
%
This distribution is naturally obtained for a population of {\em  fading} sources with intrinsic luminosity that decreases over a time
scale $\tau$ according to:
\begin{equation}
\LTeV(t)= \LMaxTeV\left(1+\frac{t}{\tau}\right)^{-\gamma}
\label{lum}
\end{equation}
where $t$ indicates the time passed since source formation.
%
In this assumption, the exponent of the luminosity distribution is given
by $\alpha = 1/\gamma + 1$.

The above description can be applied to potential TeV sources in the Galaxy, such as PWNe \citep{Gaensler:2006ua} or TeV Halos \citep{Linden:2017blp}, which are connected  with the explosion of core-collapse SN and the formation of a pulsar. 
The birth rate of these objects is similar to that of
SN explosions in our Galaxy, i.e. $R\simeq R_{\rm SN} = 0.019\,{\rm yr}^{-1}$ \cite{Diehl:2006cf}. 
If gamma-ray emission is powered by pulsar activity, the TeV-luminosity can be connected to the pulsar spin-down power, i.e.: 
\begin{equation}
\LTeV = \lambda\, \dot{E}
\end{equation}
where $\lambda \le 1$ and:
\begin{equation}
\dot{E} = \dot{E}_0 \left(1+\frac{t}{\tau_{\rm sd}}\right)^{-2}
\end{equation}
for energy loss dominated by magnetic dipole radiation (braking index $n=3$).
This implies that the fading timescale is determined by the pulsar spin-down time scale, i.e.  $\tau = \tau_{\rm sd}$. 
Moreover, if the efficiency of TeV emission does not depend on time
($\lambda\sim {\rm const}$), the exponent in Eq.~\ref{lum} is
$\gamma=2$, that corresponds to a source luminosity function $\YTeV(\LTeV)\propto \LTeV^{-1.5}$.
The possibility of $\lambda$ being correlated to the spin-down power, i.e. $\lambda = 
\lambda_0 ({\dot E}/{\dot E_0})^{\delta}$, was suggested by Abdalla {\em et al.} \cite{Abdalla:2017vci} that found $\LTeV = \lambda\,{\dot E} \propto \dot{E}^{1+\delta}$ with $1 + \delta = 0.59 \pm 0.21$ by studying a sample of PWNe in the HPGS catalog. In this case, one obtains $\gamma \simeq 1.2$ in Eq.~\ref{lum} that corresponds to a source luminosity function $\YTeV(\LTeV)\propto \LTeV^{-1.8}$.  
We consider this last scenario ($\alpha=1.8$) as our reference case, conforming to previous analyses on the subject \citep{Acero:2016qlg, Pothast:2018bvh}. 
In order to discuss thoroughly the dependence of our results on the performed assumptions, we also consider, however, the alternative hypothesis $\alpha = 1.5$.

It is finally useful to introduce the function $\mathcal{Y}_{\rm TeV}\left(\FTeV \right)$ that describes the source emissivity distribution. This is related to the luminosity function by the expression $\mathcal{Y}_{\rm TeV}\left(\FTeV \right) = \ETeV\,\YTeV(\FTeV \ETeV)$. By using Eq.~\ref{LumDist1}, we see that the emissivity distribution is not modified, if the ratio $\FMaxTeV\equiv\LMaxTeV/\ETeV$ is kept constant.

%

\subsection{Consistency among HGPS and Fermi-LAT catalogs}

\begin{figure}[htbp]
\includegraphics[width=0.45\textwidth]{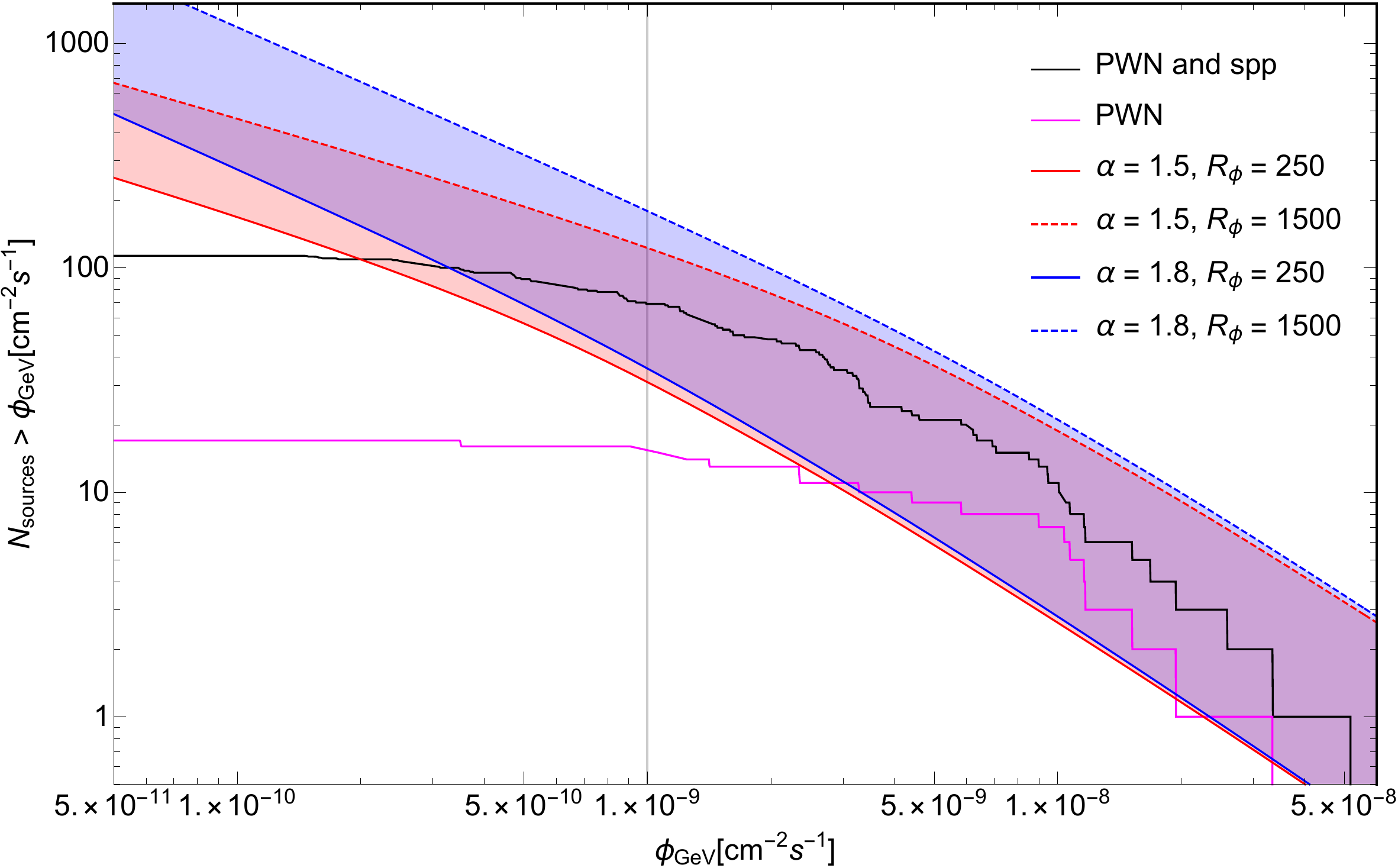}
\caption{\small\em {\bf Pulsar wind nebulae flux distribution in the GeV domain.} We report with shaded bands the cumulative number $N(\PhiGeV)$ of sources with fluxes larger than $\PhiGeV$ predicted in our model and in the latitude range $|b|\le 20.25^\circ$. The red (blue) band is obtained by assuming $\alpha=1.5$ ($\alpha=1.8$) and by considering $250\le \Rphi \le 1500$. 
The magenta line represents the cumulative number of pulsar wind nebulae (PWNe) with fluxes larger than $\PhiGeV$ in the 4FGL-DR2 catalog.   
The black line also includes SPP sources.
The gray vertical line represents the Fermi-LAT sensitivity threshold for objects in the Galactic plane. 
\label{fig:cumulatives}}
\end{figure}

%
The adopted range for the flux ratio parameter $\Rphi$ can be further validated by comparing the predicted source flux distribution in the GeV domain with the 4FGL-DR2 catalog (see Fig.~\ref{fig:cumulatives}).
It should be remarked that, while PWNe provide the prominent contribution of the observed emission at TeV energies, they are instead a subdominant component in the GeV domain. 
The 4FGL-DR2 catalog includes $5788$ sources which are mostly extragalactic objects \citep{Fermi-LAT:2019yla}.
The total number of identified and/or associated Galactic sources is $486$. 
The largest source class, including $271$ objects, is given by pulsars that typically have soft emission spectra with cut-off at few GeV and are not expected to contribute to the population of TeV emitting sources potentially detectable by HGPS. 
In addition to pulsars, the 4FGL-DR2 catalog encompasses $18$ PWNe, $43$ SNRs, and $96$ objects (labelled as SPP) of unknown nature but overlapping with known SNRs or PWNe. 
The magenta line in Fig.\ref{fig:cumulatives} corresponds to the cumulative number $N(\PhiGeV)$ of PWNe with flux larger than $\PhiGeV$ in the latitude range $|b|\le 20.25^\circ$ while the black line also includes SPP sources.
The SPP source class is not expected to fully correspond to the population considered in this work; it can be however regarded as an upper limit for theoretical calculations.

The two shaded bands in Fig.\ref{fig:cumulatives} show theoretical predictions of our population model for two different values of the power-law index of the luminosity function ($\alpha = 1.5$ and $1.8$).
Namely, the red (blue) shaded band is obtained by assuming the
best-fit values $\LMaxTeV = 4.9 \cdot 10^{35}\lUnit$ ($\LMaxTeV=6.8\cdot 10^{35}\lUnit$) and $\tau =1.8\cdot 10^3\,{\rm y}$ ($\tau = 0.5 \cdot 10^3\,{\rm y}$) for $\alpha=1.5$ ($\alpha=1.8$) given in Cataldo {\em et al.} \cite{Cataldo:2020qla} and by varying the flux ratio in the range \textbf{$250\le \Rphi \le 1500$}.
The  lower  bound for the flux ratio (i.e. $\Rphi = 250$) is obtained by requiring that
sources are not underpredicted (within statistical fluctuations) in
the flux region where the catalog can be considered complete.
More precisely, it corresponds to assuming $6$ sources with flux larger than $5\times10^{-9}\;{\rm cm}^{-2}\;{\rm s}^{-1}$ to be compared with 9 PWNe in the 4FGL-DR2 catalog.  
The upper bound (i.e. $R_\Phi=1500$) is instead obtained by requiring that very bright sources are not overpredicted and corresponds to assuming $3$ sources with flux larger than $5\times 10^{-8}\,{\rm cm}^{-2}\,{\rm s}^{-1}$ to be compared with no observed PWNe + SPP sources in the 4FGL-DR2 catalog.
%
%
%
%
%
%
%
%

In general, we see that a reasonable agreement exists with theoretical expectations, supporting the phenomenological description adopted in this paper.
%
We also note that the performed comparison provides by itself a proof that the average spectral index $\betaGeV $ of PWNe at GeV energies should be smaller than the value $\betaTeV = [2.3-2.4]$ observed in the TeV domain.
Indeed, if we assume that source spectrum is described by undistorted power-law with spectral index $\betaTeV\sim [2.3-2.4] $, we obtain $\Rphi = 10^{3\,(\betaTeV-1)}\sim 10^4$.
Considering that bright sources in the HGPS catalog have fluxes $\PhiTeV\sim10^{-11} \phiUnit$, we should expect an ensemble of sources with fluxes $\PhiGeV \sim 10^{-7}\phiUnit$ in the GeV domain. 
This is not observed by Fermi-LAT, indicating that TeV galactic sources typically have a spectral break and harder emission spectrum below $\sim1\,{\rm TeV}$.
Coherently with this conclusion, most of the PWNe in the 4FGL-DR2 catalog have a spectral indexes $\le 2$ at GeV energies.

\subsection{Total luminosity and flux}
The  total luminosity produced by the considered population in the TeV domain is given as a function of $\LMaxTeV$ and $\tau$ by: 
\begin{equation}
\LMWTeV= 
\frac{{\mathcal N}\,\LMaxTeV}{\left(2-\alpha\right)} 
\left[1 -\Delta^{\alpha-2}\right]
\label{Lmw}
\end{equation}
where ${\mathcal N} = R \, \tau \, (\alpha-1)$ and $\Delta \equiv \LMaxTeV/\LMinTeV$. Unless otherwise specified, we quote results obtained for $\Delta \to \infty$ that can be easily recalculated by using the above equation, if other values are considered.
The flux in the TeV domain produced at Earth by all sources included in a fixed observational window (OW) can be expressed as:
\begin{equation}
\PhiTotTeV = 
\xi\;
\frac{\LMWTeV} {4\pi \ETeV}\; 
\langle r^{-2} \rangle
\label{phitot}
\end{equation}
where the parameter $\xi$, which is defined as
\begin{equation}
\xi \equiv \int _{\rm OW}d^3r \, \rho({\bf r}),
\end{equation}
represents the fraction of sources of the considered population which are included in the OW, while the quantity
\begin{equation}
\langle r^{-2} \rangle \equiv \frac{1}{\xi}\int_{\rm OW}d^3r \, \rho({\bf r}) \, r^{-2} 
\end{equation}
is the average value of their inverse square distance. 
By considering Eqs.~\ref{Lmw} and \ref{phitot}, we see that the total flux produced by the considered population in the TeV domain is directly determined by the maximal emissivity $\FMaxTeV = \LMaxTeV/\ETeV$.

The total flux produced in the GeV domain can be calculated as a function of the parameter $\Rphi$ (for fixed values of $\LMaxTeV$ and $\tau$) and it is given by Eq.~\ref{Eq:PhiTotGeV}.

 \paragraph{Source flux distributions}
 The source flux distribution in the TeV domain $dN/d\PhiTeV$ can be calculated as a function of $\LMaxTeV$ and $\tau$ by using:
\begin{equation}
\frac{dN}{d\PhiTeV} = \int dr\; 
4\pi r^4
\ETeV \;
\YTeV(4\pi r^2 \ETeV \PhiTeV)\;
\overline{\rho} (r), \;
\label{dNdphi}
\end{equation}
where $\overline{\rho}(r) \equiv \int_{\rm OW} d\Omega \; \rho(r, {\bf n})$.
The above expression can be recasted in terms of the source emissivity distribution as:
\begin{equation}
\frac{dN}{d\PhiTeV} = \int dr\; 
4\pi r^4 \;
{\mathcal Y}_{\rm TeV}(4\pi r^2 \PhiTeV)\;
\overline{\rho} (r), \;
\label{dNdphi}
\end{equation}
By considering that the function ${\mathcal Y}_{\rm TeV}(\FTeV)$ only depends on the parameter $\FMaxTeV$, we can understand the effects of a variation of $\betaTeV$ in our analysis.
Indeed, a modification of $\betaTeV$ reflects into a variation of the photon average energy $\ETeV$. This can be reabsorbed by a shift of $\LMaxTeV$ in such a way that the ratio $\FMaxTeV = \LMaxTeV/\ETeV$ is kept constant, with no effects on the predicted source distribution $dN/d\PhiTeV$, on the cumulative flux $\PhiTotTeV$ in the TeV domain and on the quality of the fit to HGPS catalog.

Finally, the source flux distribution $dN/d\PhiGeV$ in the GeV domain is connected to that in TeV domain by the $\Rphi$ parameter, according to Eq.~\ref{Eq:dNdphiGeV}.
Note that the distribution $dN/d\PhiGeV$ is predicted independently on the assumed values of the spectral parameters $E_0$ and $\betaTeV$.
%
%
Faint sources that produce a flux at Earth below the Fermi-LAT observation threshold $\PhiGeVTh$ are not resolved and contribute to the large scale diffuse emission from the Galaxy. This contribution is evaluated by using Eq.~\ref{Eq:PhiGeVUnr}.
\section{Data Availability}
The data that support the findings of this study are available from the corresponding author upon reasonable request.

\section{Acknowledgement \label{sec:acknowledgement}}

We thanks Daniele Gaggero and Mart Pothast for providing us the Fermi-LAT data points of the total gamma ray diffuse emission.
The work of GP and FLV is partially supported by the research grant number 2017W4HA7S ''NAT-NET:
Neutrino and Astroparticle Theory Network'' under the program PRIN 2017 funded by the Italian Ministero dell'Istruzione, dell'Universita' e della Ricerca (MIUR).

\section{Author contributions}
All the Authors equally contributed to this work.

\section{Competing Interests}
The authors declare no competing interests.

\begin{deluxetable*}{l|c|cc|cc}
\tablenum{1}
\tablecaption{{\bf The cumulative gamma fluxes due to unresolved PWNe. }\em The cumulative flux ($\PhiGeVUNRES$) of unresolved  TeV PWNe in the GeV domain for $\alpha=1.8$ and $\alpha=1.5$ and for the two extreme values of $\Rphi$ allowed in our analysis.
In brackets, we give the percentage of unresolved sources emission with respect to the total diffuse $\gamma$-ray flux ($\Phi_{\rm GeV}^{\rm diff}$) measured by Fermi-LAT in each galactocentric ring and in the latitude window $|b| < 20.25^{\circ}$.
\label{Tab1}}
\tablewidth{0pt}
\tablehead{
\nocolhead{.} & \colhead{$\Phi_{\rm GeV}^{\rm diff}$ ($cm^{-2}\ s^{-1}$)} &
\multicolumn{2}{c}{$\PhiGeVUNRES$ ($cm^{-2}\ s^{-1}$)} & 
\multicolumn{2}{c}{$\PhiGeVUNRES$ ($cm^{-2}\ s^{-1}$)}\\
\nocolhead{.} & \nocolhead{.} &
\colhead{$\Rphi=250$, $\alpha=1.8$} & \colhead{$ \Rphi=1500$, $\alpha=1.8$} & \colhead{$ \Rphi=250$, $\alpha=1.5$} & \colhead{$ \Rphi=1500$, $\alpha=1.5$}
}
\startdata
$1.7 - 4.5\;{\rm kpc}$ & $3.86\times10^{-7}$ & $3.35\times10^{-8}\;(8.6\%)$ & $1.40\times10^{-7}\; (36\%)$  & $1.60\times10^{-8}\;(4.1\%)$ & $3.92\times10^{-8}\; (10\%)$ \\
 $ 4.5 - 5.5\;{\rm kpc}$ & $3.11\times10^{-7}$ & $1.91\times10^{-8}\;(6.1\%)$ & $8.00\times10^{-8}\;(26\%)$ & $8.30\times10^{-9}\;(2.7\%)$ & $2.00\times10^{-8}\;(6.4\%)$ \\
 $ 5.5 - 6.5\;{\rm kpc}$ & $5.09\times10^{-7}$ & $2.13\times10^{-8}\;(4.2\%)$ & $8.93\times10^{-8}\;(17\%)$ & $8.33\times10^{-9}\;(1.6\%)$ & $2.02\times10^{-8}\;(3.9\%)$ \\
 $ 6.5 - 7.0\;{\rm kpc}$ & $2.57\times10^{-7}$ & $1.15\times10^{-8}\;(4.5\%)$ & $4.81\times10^{-8}\;(19\%)$ & $3.96\times10^{-9}\;(1.5\%)$ & $9.48\times10^{-9}\;(3.7\%)$ \\
 $ 7.0 - 8.0\;{\rm kpc}$ & $7.7\times10^{-7}$ & $2.67\times10^{-8}\;(3.5\%)$ & $1.12\times10^{-7}\;(14\%)$ & $7.53\times10^{-9}\;(1.0\%)$ & $1.83\times10^{-8}\;(2.4\%)$\\
 $ 8.0 - 10.0\;{\rm kpc}$ & $3.84\times10^{-6}$ & $4.89\times10^{-8}\;(1.3\%)$ & $2.05\times10^{-7}\;(5.3\%)$ & $1.08\times10^{-8}\;(0.3\%)$ & $2.69\times10^{-8}\;(0.7\%)$\\
$ 10.0-16.5\;{\rm kpc}$ & $7.68\times10^{-7}$ & $1.51\times10^{-8}\;(1.9\%)$ & $6.37\times10^{-8}\;(8.3\%)$  & $6.37\times10^{-9}\;(0.8\%)$ & $1.65\times10^{-8}\;(2.1\%)$  \\
$ 16.5-50.0\;{\rm kpc}$ & $4.44\times10^{-8}$ & $3.87\times10^{-10}\;(0.8\%)$ & $2.07\times10^{-9}\;(4.7\%)$ & $2.43\times10^{-10}\;(0.5\%)$ & $6.98\times10^{-10}\;(1.6\%)$ \\
 \hline
 $ 0.0-50.0\;{\rm kpc}$ & $6.89\times10^{-6}$ & $1.79\times10^{-7}(2.6\%)$ & $7.53\times10^{-7}\;(11\%)$ & $6.28\times10^{-8}(1.0\%)$ & $1.54\times10^{-7}\;(2.2\%)$
\enddata
\end{deluxetable*}

\begin{table}
\tablenum{2}
\caption{\em {\bf Gamma Ray Spectral Indexes.} Spectral indexes of the CR diffuse emission obtained by fitting the Fermi-LAT data 
with ($\Gamma_{BF}$) and without ($\Gamma_1$) TeV PWNe unresolved contribution. The first error associated to $\Gamma_{BF}$ represents the systematic uncertainty (due to variations of $\Rphi$, $E_{0}$ and $\betaTeV$) while the second is the statistical one.
The indexes $\Gamma_1$ coincide with those obtained by Pothast {\em et al.} \cite{Pothast:2018bvh}.}
\label{tab:gamma18}
\hspace{-1.6cm}
\begin{tabular}{l|c|l|l }
\hline
\hline
Ring {$({\rm kpc})$} & $ \Gamma_1$ &  $\Gamma_{BF}$ ($\alpha=1.8$) &  $\Gamma_{BF}$ ($\alpha=1.5$)\\
 \hline
$1.7-4.5$ & $2.56 \pm 0.02$ & $2.71^{+0.19}_{-0.09}\pm0.01$ & $2.60^{+0.10}_{-0.03}\pm0.01$  \\
$4.5 - 5.5$ & $2.48\pm0.02$ & $2.56^{+0.11}_{-0.05}\pm0.01$ & $2.50^{+0.06}_{-0.02}\pm0.01$ \\
$5.5 - 6.5$ & $2.53\pm0.02$ & $2.62^{+0.10}_{-0.04}\pm0.01$ & $2.57^{+0.05}_{-0.01}\pm0.01$ \\
$6.5 - 7$ & $2.52\pm0.02$ & $2.62^{+0.10}_{-0.05}\pm0.01$ & $2.56^{+0.05}_{-0.01}\pm0.01$  \\
$7 - 8$ & $2.58\pm0.01$ & $2.62^{+0.07}_{-0.03}\pm0.008$ & $2.58^{+0.02}_{-0.008}\pm0.008$ \\
$8 - 10$ & $2.64\pm0.01$ & $2.66^{+0.03}_{-0.01}\pm0.004$ & $2.64^{+0.01}_{-0.004}\pm0.004$ \\
$10 - 16.5$ & $2.68\pm0.02$ & $2.74^{+0.05}_{-0.03}\pm0.009$ & $2.70^{+0.04}_{-0.008}\pm0.008$ \\
$16.5 - 50$ & $2.73\pm0.05$  & $2.77^{+0.10}_{-0.04}\pm0.04$ & $2.73^{+0.08}_{-0.03}\pm0.04$\\
\hline
\hline
\end{tabular}
\end{table}

\newpage

\bibliography{bibliography.bib}
\bibliographystyle{aasjournal}

\appendix

\section{Supplementary Materials}


%

\paragraph{PWNe in 4FGL-DR2 and HGPS catalogs}

In Tab.\ref{tab:pwn}, we discuss the spectral properties of sources that are firmly identified both in the 4FGL-DR2 and HGPS catalogs.
We also consider the CRAB nebula that is well studied at TeV
energy (even if it is not included in the HGPS catalog). 
%
%
Namely, we show the value of the flux ratio $R_{\Phi}$, the spectral indexes in the GeV and TeV domain, $\beta_{\rm GeV}$ and $\beta_{\rm TeV}$, the source distance $D$ (if available) and its characteristic age~\citep{Giacinti:2019nbu}.
We see that the average values for $R_{\Phi}$ and $\beta_{\rm GeV}$ are $1122$ and $1.89$, respectively. 
The average age of the considered PWNe population $11.7$~kyr.
This additional information can be useful to compare with theoretical predictions of the SED produced by Inverse Compton emission, see for example the expected emission for young PWNe given in Fig.~A1 of \citep{Abdalla:2017vci}.

\begin{table}
\caption{\em {\bf PWNe observed both in H.E.S.S. and in Fermi-LAT:} The $12$ PWNe included both in 4FGL-DR2 and HGPS catalogs (with the addition of  CRAB \citep{HESS:2006fka}, although not included in HGPS). 
In the first column, we give the source name in HGPS catalog.
In the second column, we show the flux ratio $R_{\Phi} = \Phi_{\rm GeV}/\Phi_{\rm TeV}$, obtained by considering the 4FGL-DR2 and HGPS  observational determinations of  $\Phi_{\rm GeV}$ and $\Phi_{\rm GeV}$, respectively. 
In the third and fourth columns, we report the power-law spectral indexes $\beta_{\rm GeV}$ and $\beta_{\rm TeV}$ in the GeV and TeV energy domains. 
The fourth and fifth columns give the source distance and characteristic age.}

\begin{tabular}{l|c|c|c|c|c}

\hline
\hline
 H.E.S.S.-association &
$R_{\Phi}$   & $\beta_{\rm GeV}$ & $\beta_{\rm TeV}$ &  D(kpc) &  $\tau_{c}$(kyr) \\
\hline
CRAB & $1481$ & $1.38$ (1 GeV) (log-par) & $2.39$ & $2.0$ & $0.94$ \\
HESS J0835-455 & $754$ & $2.18$ & $1.89$ & $0.29$ & $11.3$ \\
HESS J1303-631 & $447$ & $1.81$ & $2.33$ & $6.7$ & $11.0$ \\
HESS J1356-645 & $63$ & $1.41$ & $2.20$  & $2.4$ & $7.31$    \\
HESS J1420-607 & $999$ & $1.99$ & $2.20$ & $5.6$ & $13.0$    \\
HESS J1514-591 & $686$ & $1.83$ & $2.05$ & $5.2$   & $1.56$\\  
HESS J1616-508 & $1223$ & $2.05$ & $2.32$ & $6.8$ & $8.13$  \\
HESS J1632-478 & $799$ & $1.76$ & $2.51$ & - & - \\
HESS J1746-285$^{*}$ & $98950$ & $0.96$ (1 GeV) (log-par) & $2.17$ & - & -  \\
HESS J1825-137 & $582$ & $1.73$ & $2.38$ & $3.9$ & $21.4$ \\
HESS J1837-069 & $1612\;(483)$ & $2.04\;(1.84)$ & $2.54$ & $6.6$  & $22.7$  \\
HESS J1841-055 & $1149$ & $1.98$ & $2.47$ & -  & - \\
HESS J1857+026 & $2390$ & $2.12$ & $2.57$ & - & $20.6$ \\ 
\hline
\hline
\end{tabular}
\\
\\
$^*$ \textit{This source shows unexpected energy cutoff in the Fermi-LAT spectrum apparently not compatible with its TeV counterpart.}
\label{tab:pwn}
\end{table}


\begin{figure}[h!]
\begin{center}
\includegraphics[width=0.5\textwidth]{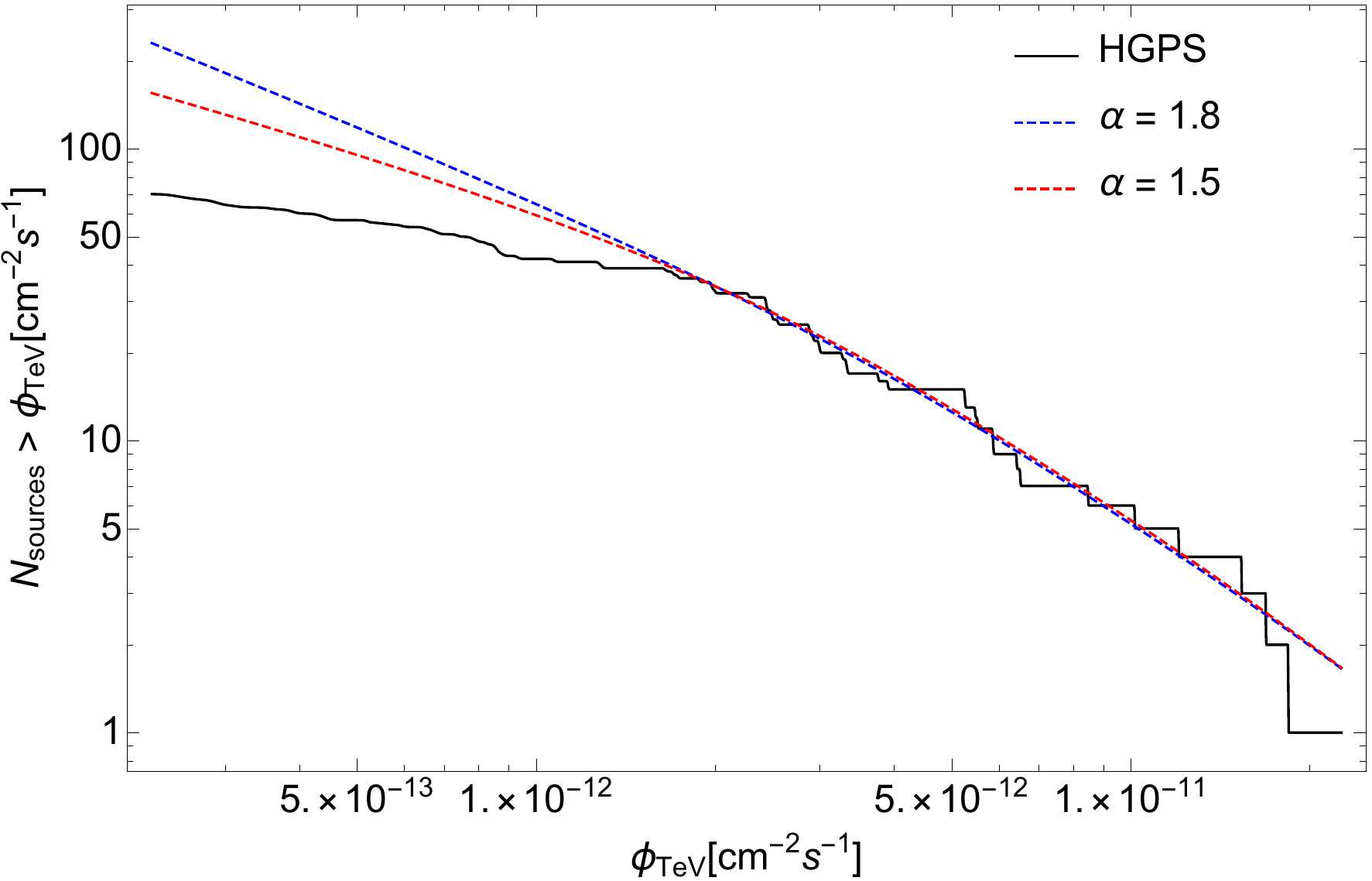}\\
\caption{\small\em {\bf Comparison between observational and
theoretical cumulative number of sources in the TeV energy range:} The cumulative distribution of source in HGPS is shown by a black thick line. The blue (red) dashed line represent theoretical predictions from our population model for the best fit values of the maximal luminosity $\LMaxTeV=6.8\cdot 10^{35}{\rm \;erg\;s^{-1}}$ ($\LMaxTeV=5.0\cdot 10^{35}{\rm \;erg\;s^{-1}}$) and spin-down timescale $\tau = 0.5 \cdot 10^3\,{\rm y} $ ($\tau = 1.7 \cdot 10^3\,{\rm y} $) with luminosity index $\alpha=1.8$ ($\alpha=1.5$).
\label{fig:cumHess}}
\end{center}
\end{figure}

\paragraph{Comparison with HGPS catalog}

In Fig.~\ref{fig:cumHess}, we compare theoretical predictions from our population model
with the cumulative distribution of HGPS sources (black solid line).
The theoretical distribution for $\alpha=1.8$ ($\alpha=1.5$) is shown by a blue dashed line (red dashed line) and it is calculated by assuming that the maximal PWNe luminosity and spin-down timescale are $\LMaxTeV =6.8\cdot 10^{35}{\rm \;erg\;s^{-1}}$ ($\LMaxTeV =5.0\cdot 10^{35}{\rm \;erg\;s^{-1}}$) and   $\tau = 0.5 \cdot 10^3\,{\rm y} $ ($\tau = 1.7 \cdot 10^3\,{\rm y} $), respectively.
These values have been obtained by performing an unbinned likelihood fit of the flux, latitude and longitude distribution of bright sources in the HGPS catalog (and assuming that the PWNe birth rate is equal to that of core-collapse SN explosions, i.e.  $R=0.019\;{\rm yr}^{-1}$).\\

\paragraph{The effects of assumed spectral parameters on unresolved PWNe emission}

\begin{figure}[b!]
\begin{center}
\includegraphics[width=0.55\textwidth]{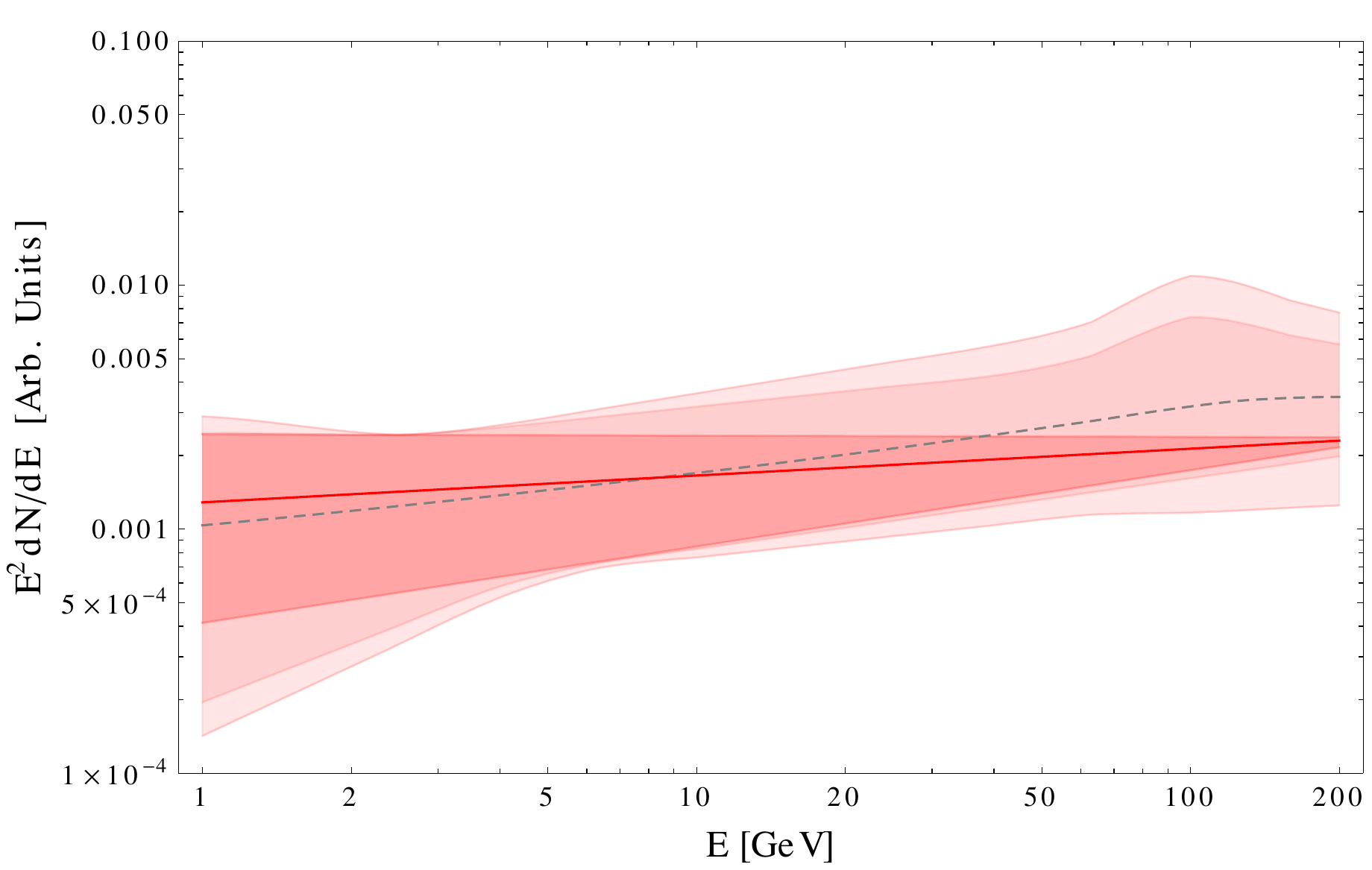}\\
\caption{\small\em {\bf Unresolved PWN spectrum:} The effects of assumed spectral parameters on the unresolved PWNe emission. The shaded bands are obtained by varying only $\Rphi$ (red), by varying $\Rphi$ and $E_0$ (light red) and by considering simultaneous variations of $\Rphi$, $E_0$ and $\betaTeV$ (pink). See text for details. 
\label{fig:SpettriPWNe_Irr}}
\end{center}
\end{figure}

 A modification of the spectral parameters $\Rphi$, $E_0$ and $\betaTeV$ reflects into a fractional variation of the unresolved PWNe emission that is identical in each ring. 
The produced effects are shown in Fig.~\ref{fig:SpettriPWNe_Irr}.
Here, the red line corresponds to predicted emission for the case $\Rphi \simeq 770$, $\betaTeV\simeq 2.4$ and $E_0 =0.8$~TeV that well reproduces the cumulative spectral energy distribution of sources observed both by Fermi-LAT and H.E.S.S. (see Sect.~\nameref{sec:method})
The shaded bands are obtained by varying only $\Rphi$ (red), by varying $\Rphi$ and $E_0$ (light red) and by considering simultaneous variations of $\Rphi$, $E_0$ and $\betaTeV$ (pink).
To be quantitative, we note that the unresolved PWNe emission at $50$~GeV can be increased (decreased) by a factor $\sim 3$ ($\sim 2$) with respect to the reference case when $(R_{\Phi}, \, E_{0} , \, \beta_{\rm TeV})$ are simultaneously varied in the 3-dim parameter space defined by the ranges 
$R_\Phi= \left[250-1500\right]$,  $E_0 = \left[0.1-1.0 \right] \, {\rm TeV} $  and $\beta_{\rm TeV}=\left[1.9 -2.5 \right]$.
For completeness, we also show with a dashed line the predicted emission that was used to obtain the central values for $\Gamma_{BF}$ quoted in Tab.~\ref{tab:gamma18}.
This is obtained by integrating over the whole parameters space.
We assume logarithmic uniform distributions for the spectral break position
and for the flux ratio,
while for $\betaTeV$ we consider a Gaussian distribution centered in $\betaTeV=2.4$ and with dispersion $0.15$ as reported in the HGPS catalog~\citep{H.E.S.S.:2018zkf}.%

\end{document}